\documentclass[%
 reprint,
superscriptaddress,
nofootinbib,
 amsmath,amssymb,
 aps,
longbibliography,
showkeys,
prb,
]{revtex4-2}
\usepackage{hyperref}
\usepackage{amsmath}
\usepackage{amssymb}
\usepackage{cleveref}
\usepackage{appendix}
\usepackage{braket}
\usepackage{bbold}
\usepackage[english]{babel}
\usepackage{natbib}
\usepackage{xcolor}
\usepackage{graphicx}
\usepackage{siunitx}

\usepackage{comment}

\begin{document}

\title{Helicity of Stationary Chiral Molecules}

\author{Ivanina Ilieva}
\email{ivanina.ilieva@kit.edu}
\affiliation{Institute of Theoretical Solid State Physics, Karlsruhe Institute of Technology, Kaiserstr. 12, 76131 Karlsruhe, Germany}
\author{Christof Holzer}
\affiliation{Institute of Quantum Materials and Technologies, Karlsruhe Institute of Technology, Kaiserstr. 12, 76131 Karlsruhe, Germany}
\author{Lukas Rebholz}
\affiliation{Institute of Theoretical Solid State Physics, Karlsruhe Institute of Technology, Kaiserstr. 12, 76131 Karlsruhe, Germany}
\author{Carsten Rockstuhl}
\affiliation{Institute of Theoretical Solid State Physics, Karlsruhe Institute of Technology, Kaiserstr. 12, 76131 Karlsruhe, Germany}
\affiliation{Institute of Nanotechnology, Karlsruhe Institute of Technology, Kaiserstr. 12, 76131 Karlsruhe, Germany}
\affiliation{Center for Integrated Quantum Science and Technology (IQST), Germany}
\author{Ivan Fernandez-Corbaton}
\affiliation{Institute of Nanotechnology, Karlsruhe Institute of Technology, Kaiserstr. 12, 76131 Karlsruhe, Germany}

\date{November 27, 2026}

\begin{abstract}
The helicity of material systems can be considered to be the same fundamental quantity as the helicity of light. 
With chirality being a prerequisite for non-zero helicity, twisted magnetization densities were thought to be the only kind of sources able to store helicity in static matter. 
In this work, we extend the types of electromagnetic sources in the stationary regime that can store helicity and study the helicity contained in stationary chiral molecules in the form of non-trivial magnetization and {\em transverse current} density configurations. 
The consideration of the latter is highly important for the study of the helicity of molecules, as in some cases it is the only quantity capable of storing helicity. 
We develop the theoretical formalism needed to compute helicity from a given set of stationary magnetization and current densities, and then apply it to two examples simulated using non-collinear spin density functional theory: the organic molecule BINOL under the influence of an external magnetic field and an open-shell, Europium-based single-molecule magnet. 
While the helicity of the Europium-compound is contained in both its magnetization and current densities, the helicity in BINOL comes only from its stationary current density, since its magnetization density is identically zero. 
We compute the helicity stored in each of the molecules, and give an outlook on the importance of the stored helicity for chiral light--matter interactions.
\end{abstract}

\keywords{helicity, chirality, chiral molecules, magnetic helicity}

\maketitle
\section{Introduction and Summary}
\label{sec:introduction}
The helicity of a vector field represents the topological properties of its field lines -- the degree of linkage and knottedness, and the sense of twist and writhe \cite{moreauConstantesDunIlot1961,moffattDegreeKnottednessTangled1969,bergerTopologicalPropertiesMagnetic1984}. 
It is related to a topological invariant called the Hopf charge, which takes integer values corresponding to the linking number of the vector field lines\footnote{We note that there are recent works exploring the possibility of the Hopf charge also taking non-integer values \cite{guslienkoEmergentMagneticField2023,azhar3DMagneticTextures2026}.} \cite{hopfUberAbbildungenDreidimensionalen1931,whiteheadExpressionHopfsInvariant1947,kuznetsovTopologicalMeaningCanonical1980}. 
The Hopf charge for a given vector field configuration is therefore conserved for continuous deformations of the field. 
Explicitly, the helicity of a vector field $\mathbf{w}(\mathbf{r})$ with $\mathbf{w}(\mathbf{r}) = \nabla\times \mathbf{v}(\mathbf{r})$ inside a volume $V$ is given by (see Ref.~\cite{moffattHelicity2017} and references therein)
\begin{equation}
\label{eq:helicity_of_general_vector_field}
    \int _V \mathrm{d}^3r \, \mathbf{w}(\mathbf{r})\cdot\mathbf{v}(\mathbf{r}) \, .
\end{equation}
The above expression is gauge invariant for a divergence-free field $\mathbf{w}(\mathbf{r})$ with $\mathbf{w}(\mathbf{r})\cdot \mathbf{n}(\mathbf{r})=0$ on the boundary of the volume \cite{finnMagneticHelicityWhat1985,bergerTopologicalPropertiesMagnetic1984}. 
It is relevant in different research areas. 
For example, in fluid dynamics and atmospheric physics, $\mathbf{v}(\mathbf{r})$ is a velocity field and $\mathbf{w}(\mathbf{r})$ is its vorticity, and their scalar product describes the spinning motion of the fluid along its flow direction, in which case we speak of {\it vortex} or {\it kinetic helicity} \cite{kuzanyanHelicitiesGeophysicsAstrophysics2024}. Vortex helicity is a global invariant of ideal fluid flows \cite{moreauConstantesDunIlot1961,moffattDegreeKnottednessTangled1969,moffattHelicityLaminarTurbulent1992,soltanitehraniGalileanInvarianceMean2023}. 

In plasma physics, magnetohydrodynamics, and astrophysics, $\mathbf{w}(\mathbf{r})$ is the magnetic field $\mathbf{B}(\mathbf{r})$, $\mathbf{v}(\mathbf{r})$ is the magnetic vector potential $\mathbf{A}(\mathbf{r})$, and the quantity in Eq.~(\ref{eq:helicity_of_general_vector_field}) is called {\it magnetic helicity} \cite{elsasserHydromagneticDynamoTheory1956,woltjerTheoremForcefreeMagnetic1958,moffattDegreeKnottednessTangled1969,brownMagneticHelicitySpace2010}. 
Magnetic helicity quantifies the topology of the magnetic field lines in a system and is an important tool for understanding astrophysical dynamo processes (see Refs.~\cite{boozerMagneticHelicityDynamos1993,brandenburgCriticalRoleMagnetic2009,blackmanMagneticHelicityLarge2015} and references therein), magnetic relaxation processes \cite{jiConservationMagneticHelicity1995,bellanFundamentalsPlasmaPhysics2006,yeatesEvolutionFieldLine2021} and solar coronae \cite{bergerRigorousNewLimits1984,ImplicationsMagneticHelicity,kusanoStudyMagneticHelicity2004}. 
While energy dissipates easily, magnetic helicity is conserved in systems with infinite conductivity, such as ideal plasmas \cite{woltjerTheoremForcefreeMagnetic1958}, and approximately conserved in real plasmas with finite conductivity on relevant time scales \cite{taylorRelaxationToroidalPlasma1974,bergerRigorousNewLimits1984}. 
Magnetic helicity, or, equivalently, the Hopf charge, is central in the study of topological textures in magnets, which, in turn, are relevant for the emerging field of spintronics \cite{kosevichMagneticSolitons1990,sutcliffeSkyrmionKnotsFrustrated2017,sutcliffeHopfionsChiralMagnets2018,kentCreationObservationHopfions2021,knapmanNumericalCalculationHopf2025,shajuHopfionsScrewChiral2026}. 

In electrodynamics, we have the sum of two analogous terms making up the {\it electromagnetic helicity}, one being the magnetic helicity and the other the electric helicity, where $\mathbf{w}(\mathbf{r})$ is the electric field $\mathbf{E}(\mathbf{r})$ and $\mathbf{v}(\mathbf{r})$ is the electric vector potential $\mathbf{C}(\mathbf{r})$. 
While we typically use the scalar potential $\phi (\mathbf{r})$ to express the electric field via $\mathbf{E}(\mathbf{r}) = - \nabla \phi (\mathbf{r})$, for charge-free regions, an electric vector potential can be defined in analogy with the magnetic one such that $\mathbf{E}(\mathbf{r}) = \nabla \times \mathbf{C}(\mathbf{r})$ \cite{strattonElectromagneticTheory2007}. 
For plane waves, the helicity corresponds to the circular polarization handedness of the wave, that is, the sense of twist of the direction of the electric and magnetic fields as the plane wave propagates. 
For photons, the electromagnetic helicity is the same as the helicity used in particle physics, namely, the projection of angular momentum onto the direction of linear momentum \cite{truebaElectromagneticHelicity1996}. 
From that perspective, the electromagnetic helicity of a given free electromagnetic field represents the difference between the number of left- and right-handed photons in the field and is a conserved quantity \cite{lipkinExistenceNewConservation1964,calkinInvariancePropertyFree1965,deserDualityTransformationsAbelian1976,bialynicki-birulaNoteHelicity1981,afanasievHelicityFreeElectromagnetic1996,drummondDualSymmetricLagrangians1999,fernandez-corbatonConformallyInvariantDerivation2019}. 
As such, it is a highly relevant quantity for chiral light--matter interactions \cite{tangOpticalChiralityIts2010,fernandez-corbatonHelicityAngularMomentum2012, liningerChiralityLightMatter2023,poulikakos_OpticalChiralityFlux2016,cameronChiralityAngularMomentum2017,poulikakos_optChirality_inhomSpace,poulikakosChiralLightMatterInteractions2018,poulikakosOpticalHelicityOptical2019}.

In this work, we apply the concept of helicity to a new system: stationary chiral molecules. 
Chiral molecules lack inversion symmetry, which is a necessary condition for a vector field to have non-zero helicity, as follows directly from Eq.~(\ref{eq:helicity_of_general_vector_field}).  
For the description of our systems, we extend the theoretical framework introduced in Refs.~\cite{fernandez-corbatonTotalHelicityElectromagnetic2021,fernandez-corbatonScalarProductComputing2023}, where, in matter, helicity could only be contained in twisted static magnetization densities. In here, we consider that in the stationary regime, molecules can have both a magnetization density coming from electron spin as well as a current density coming from spin-orbit coupling or arising as a result of an applied external magnetic field. 
These quantities are chiral if the molecule itself is chiral and, hence, allow for non-zero helicity stored in the molecule. 
We simulate the molecular systems in this work and numerically acquire the needed quantities using non-collinear spin density functional theory (sDFT) with the software {TURBOMOLE} \cite{franzkeTURBOMOLETodayTomorrow2023}. 
From the resulting magnetization and current densities, we calculate the helicity stored in different chiral molecules. 
We study the helicity stored in BINOL, a chiral organic molecule, which, when placed in an external magnetic field, exhibits a current density. 
Its magnetization density vanishes identically, making the consideration of the current density be of utmost importance. 
We explore how the helicity changes with the strength of the external magnetic field and find a quadratic dependence. 
The spatial distribution of the helicity integrand is visualized and shows that it conforms to the remaining $C_2$ rotational symmetry of the system. 
Then, a bigger and more complicated Europium-based open-shell molecule is analyzed, which has both helicity storage channels in the stationary regime -- magnetization and current density. 
We calculate the total helicity and break it down to the individual contributing terms to find that the interplay between the current and magnetization densities yields the most important contribution in this case. 
We note that the formalism lends itself to all system sizes and is not limited only to molecules. 
The helicity stored in chiral material systems could influence the interaction between such systems and light; the topic is left as an outlook beyond the current article. 

This article is structured as follows. 
In Sec.~\ref{sec:theory}, we lay down the theoretical foundation for calculating the helicity stored in a stationary molecule. 
We give an expression for the helicity stored in the electric and magnetic fields of a stationary system. 
Then, starting from Maxwell's equations, we provide expressions using directly the quantities that allow for such storage, namely, the stationary magnetization and current densities. 
The formulas are given both in $\mathbf{k}$-space and real space. 
Finally, we validate our theoretical results by showing that they are analogous to a derivation of helicity stored in magnetization and current densities, based on the magnetic vector potential formulation of helicity, the magnetic helicity. 
In Sec.~\ref{sec:results}, we present the numerical results. 
Starting with Subsec.~\ref{subsec:methods}, we give an overview of the numerical methods used to simulate the molecular systems and retrieve the needed quantities. 
In Subsec.~\ref{subsec:binol}, we discuss the numerical results for BINOL and in Subsec.~\ref{subsec:europium_magnet}, those for the Europium-based open-shell molecule. 
We lay out the conclusions in Sec.~\ref{sec:conclusions}.

\section{Theoretical Results}
\label{sec:theory}
In the following, we introduce the theoretical framework used in this work. 
We adopt the formalism presented in Refs.~\cite{fernandez-corbatonTotalHelicityElectromagnetic2021,fernandez-corbatonScalarProductComputing2023}, which describes {\it static material systems} and allows us to calculate quantities, one being helicity, stored in such systems. 
Here, the quantity we are interested in is helicity, and the systems -- {\it stationary chiral molecules}. 
Therefore, in this section, we present a brief summary of the existing theoretical formalism and then extend it. 
Using the Dirac braket notation, the state ket describing the material system can be taken as
\begin{equation}
    \label{eq:ket_phi_E_B}
    \ket{\Phi} \equiv \begin{pmatrix}
        \sqrt{\varepsilon_0}~\overline{\mathbf{E}} (\mathbf{k})\\ \frac{\overline{\mathbf{B}}(\mathbf{k})}{\sqrt{\mu_0}}
    \end{pmatrix}
\end{equation}
with 
\begin{equation}
\label{eq:3D_FT_static_fields}
    \overline{\mathbf{E}}(\mathbf{k}) = \frac{1}{\sqrt{(2\pi)^3}} \int \mathrm{d}^3 r \,\mathbf{E}(\mathbf{r}) \text{e}^{-\text{i}\mathbf{k}\cdot\mathbf{r}}     
\end{equation}
and the equivalent expression for the magnetic field being the three-dimensional Fourier transforms of the electric and magnetic field, $\mathbf{E}(\mathbf{r})$ and $\mathbf{B}(\mathbf{r})$, respectively, with a dependence on the wave vector $\mathbf{k}$; $\varepsilon_0$ and $\mu_0$ are, correspondingly, the vacuum permittivity and permeability. 
The scalar product between two states $\ket{\Phi_1}$ and $\ket{\Phi_2}$ is given as
\begin{equation}
\label{eq:scalar_product_free_EB}
\begin{aligned}
    \braket{\Phi_1|\Phi_2} &= \int \frac{\mathrm{d}^3 k}{\hbar c |\mathbf{k}|} \begin{pmatrix}
        \sqrt{\varepsilon_0}~\overline{\mathbf{E}}_1 (\mathbf{k})\\ \frac{\overline{\mathbf{B}}_1(\mathbf{k})}{\sqrt{\mu_0}}
    \end{pmatrix} ^\dagger \begin{pmatrix}
        \sqrt{\varepsilon_0}~\overline{\mathbf{E}}_2 (\mathbf{k})\\ \frac{\overline{\mathbf{B}}_2(\mathbf{k})}{\sqrt{\mu_0}}
    \end{pmatrix} \\
    &=  \int \frac{\mathrm{d}^3 k}{\hbar c |\mathbf{k}|} \, 
       \big(\varepsilon_0 \overline{\mathbf{E}}_1 ^\dagger (\mathbf{k}) \overline{\mathbf{E}}_2 (\mathbf{k}) + \frac{1}{\mu_0} \overline{\mathbf{B}}_1 ^\dagger (\mathbf{k}) \overline{\mathbf{B}}_2 (\mathbf{k})\big) \, 
\end{aligned}
\end{equation}
with $\hbar$ being the reduced Planck constant and $c$ the speed of light. The scalar product of a state with itself, $\braket{\Phi|\Phi}$, is its norm squared. 
To express the expectation value of a given observable, e.g., energy, linear or angular momentum, etc., in a given state $\ket{\Phi}$, we let its corresponding operator act between the bra and the ket of that state.
With the helicity operator $\Lambda$ in the $\mathbf{k}$-space representation \cite{fernandez-corbatonConformallyInvariantDerivation2019}
\begin{equation}
    \label{eq:helicity_operator_k}
    \Lambda = \text{i}\hbar\hat{\mathbf{k}}\times \, ,
\end{equation}
where $\hat{\mathbf{k}}=\frac{\mathbf{k}}{|\mathbf{k}|}$, the expectation value of the helicity of a system represented by the state $\ket{\Phi}$ is
\begin{widetext}
\begin{equation}
    \label{eq:helicity_generalform_E_B}
    \begin{aligned}
            \braket{\Lambda} &= \braket{\Phi|\Lambda|\Phi}=  \int \frac{\mathrm{d}^3 k}{\hbar c |\mathbf{k}|} \begin{pmatrix}
        \sqrt{\varepsilon_0}~\overline{\mathbf{E}} (\mathbf{k})\\ \frac{\overline{\mathbf{B}}(\mathbf{k})}{\sqrt{\mu_0}}
			\end{pmatrix} ^\dagger \begin{pmatrix}\text{i}\hbar\hat{\mathbf{k}}\times&\mathbf{0}_{3\times 3}\\\mathbf{0}_{3\times 3}&\text{i}\hbar\hat{\mathbf{k}}\times\end{pmatrix}\begin{pmatrix}
        \sqrt{\varepsilon_0}~\overline{\mathbf{E}} (\mathbf{k})\\ \frac{\overline{\mathbf{B}}(\mathbf{k})}{\sqrt{\mu_0}}
    \end{pmatrix} \,,
    \end{aligned}
\end{equation}
\end{widetext}
which we shall refer to as {\it the helicity stored in the system}. In the above equation, the helicity has dimensions of angular momentum, or, equivalently, $\hbar$. The null $3\times 3$ blocks in Eq.~\ref{eq:helicity_generalform_E_B} represent an important difference for helicity between the static and dynamic cases: the helicity operator does not mix electric and magnetic fields \cite{fernandez-corbatonTotalHelicityElectromagnetic2021}.

Equation~(\ref{eq:helicity_generalform_E_B}) provides a formula for calculating the helicity stored in a given field configuration. 
However, the fields in a stationary system are generated by the sources in it. 
Perhaps the most intuitive example is an unchanging charge distribution, which produces a static electric field. 
The specifics of the electric field can be calculated using Maxwell's equations, which provide us with the exact connection. 
The same applies to all sources that enter Maxwell's equations. 
However, we can circumvent the step of calculating the fields generated by a given set of sources and find an equivalent of Eq.~(\ref{eq:helicity_generalform_E_B}) that includes only the sources and not the fields. 
This allows us to readily evaluate the helicity stored in a system if we know the stationary sources in it.
To that end, we consider Maxwell's equations in the dynamic regime
\begin{subequations}
\label{eq:Maxwell_full}
\begin{align}
        \nabla \cdot \mathbf{E}(\mathbf{r},t) &= \frac{\rho(\mathbf{r},t)}{\varepsilon _0}  - \frac{\nabla\cdot\mathbf{P}(\mathbf{r},t)}{\varepsilon _0} \, , \label{eq:ME_full_divE} \\ \nabla \cdot \mathbf{B}(\mathbf{r},t) &= 0 \, , \label{eq:ME_full_divB} \\
        \nabla \times \mathbf{E}(\mathbf{r},t) &= - \frac{\partial\mathbf{B}(\mathbf{r},t)}{\partial t} \, , \label{eq:ME_full_curlE} \\ \nabla \times \mathbf{B}(\mathbf{r},t) &= \varepsilon_0 \mu _0 \frac{\partial\mathbf{E}(\mathbf{r},t)}{\partial t} + \mu _0 \frac{\partial\mathbf{P}(\mathbf{r},t)}{\partial t} \label{eq:ME_full_curlB} \\ & \ \ \ + \mu _0 \nabla\times\mathbf{M}(\mathbf{r},t) + \mu _0 \mathbf{J}(\mathbf{r},t) \, , \nonumber
\end{align}
\end{subequations}
where $\rho(\mathbf{r},t)$ is the charge density, which might vary in space and time, the latter of which leads to a current density $\mathbf{J}(\mathbf{r},t)$. 
Importantly, the magnetization density $\mathbf{M}(\mathbf{r},t)$ arises from the spin of the electrons in the system and, upon temporal change, leads to a polarization density $\mathbf{P}(\mathbf{r},t)$. 
We do not consider any polarization coming from the electric charges, as it would not contribute to the helicity in the stationary limit. 

In the next step, we take the stationary limit of Maxwell's equations, as the formalism will be applied to stationary systems. 
For all quantities in Eq.~(\ref{eq:Maxwell_full}), this means that we drop the time dependence and all temporal derivatives vanish. 
Furthermore, the polarization density that is due to a changing magnetization density is set to zero. 
Following the same logic and setting the current density to zero recovers the static limit discussed in Refs.~\cite{fernandez-corbatonTotalHelicityElectromagnetic2021,fernandez-corbatonScalarProductComputing2023}. 
In this work, however, we allow for stationary current densities $\mathbf{J}(\mathbf{r})$ with no divergence, $\nabla \cdot \mathbf{J}(\mathbf{r}) = 0$. 
The condition of zero divergence, $\nabla \cdot \mathbf{J}(\mathbf{r}) = 0$, ensures that the current is stationary as per the charge continuity equation 
\begin{equation}
\label{eq:continuity}
    \nabla \cdot \mathbf{J}(\mathbf{r}) + \frac{\partial}{\partial t} \rho (\mathbf{r}) = 0
\end{equation}
since we may not have a time dependence in this regime. 
With these considerations, Maxwell's equations in the stationary limit read as
\begin{subequations}
\label{eq:Maxwell_stationary_r}
\begin{align}
        \nabla \cdot \mathbf{E}(\mathbf{r}) &= \frac{\rho(\mathbf{r})}{\varepsilon _0} \, ,\label{eq:ME_stat_divE} \\ \nabla \cdot \mathbf{B}(\mathbf{r}) &= 0 \, , \label{eq:ME_stat_divB} \\
        \nabla \times \mathbf{E}(\mathbf{r}) &= 0 \, , \label{eq:ME_stat_curlE} \\ \nabla \times \mathbf{B}(\mathbf{r}) &=  \mu _0 \nabla\times\mathbf{M}(\mathbf{r}) + \mu _0 \mathbf{J}(\mathbf{r}) \label{eq:ME_stat_curlB} \, .
\end{align}
\end{subequations}
Finally, in order to rewrite Eq.~(\ref{eq:helicity_generalform_E_B}), we bring Eq.~(\ref{eq:Maxwell_stationary_r}) into its equivalent in $\mathbf{k}$-space. With the substitution $\nabla \rightarrow \text{i}\mathbf{k}$, the stationary Maxwell's equations become
\begin{subequations}
    \label{eq:Maxwell_stationary_k}
    \begin{align}
    \text{i}\mathbf{k} \cdot \overline{\mathbf{E}} (\mathbf{k}) &= \frac{\overline{\rho}(\mathbf{k})}{\varepsilon_0} \, ,  \label{eq:ME_k_divE} \\ \text{i}\mathbf{k} \cdot \overline{\mathbf{B}} (\mathbf{k}) &= 0 \, , \label{eq:ME_k_divB} \\ 
    \text{i}\mathbf{k} \times \overline{\mathbf{E}} (\mathbf{k}) &= 0 \, , \label{eq:ME_k_curlE} \\ 
    \text{i}\mathbf{k} \times \overline{\mathbf{B}} (\mathbf{k}) &= \mu_0\text{i}\mathbf{k} \times \overline{\mathbf{M}} (\mathbf{k}) + \mu _0\overline{\mathbf{J}} (\mathbf{k}) \, . \label{eq:ME_k_curlB}
\end{align}
\end{subequations}
The above equations provide significant insight into the possibility of storing helicity in stationary fields. 
Specifically, Eqs.~(\ref{eq:ME_k_curlE}) and~(\ref{eq:ME_k_curlB}) are, in essence, the helicity operator, Eq.~(\ref{eq:helicity_operator_k}), applied on the fields. 
With an eye on Eq.~(\ref{eq:helicity_generalform_E_B}), Eq.~(\ref{eq:ME_k_curlE}) tells us that there is no helicity stored in a static electric field, as the helicity operator applied on such a field yields zero. 
Together with Eq.~(\ref{eq:ME_k_divE}), we see that a stationary electric field is purely longitudinal (parallel to $\mathbf{k}$). 
The magnetic field, on the other hand, is purely transverse, as seen from Eqs.~(\ref{eq:ME_k_divB}) and~(\ref{eq:ME_k_curlB}). 
Equation~(\ref{eq:ME_k_curlB}) shows that a stationary magnetic field can, in fact, store helicity; namely, when it is coming from a magnetization or current density. 
For a magnetization density arising from spin, only the transverse (perpendicular to $\mathbf{k}$) part contributes to the helicity, as per the first term on the right-hand side of Eq.~(\ref{eq:ME_k_curlB}). Since the other two terms are obviously transverse, the second term on the right-hand side of Eq.~(\ref{eq:ME_k_curlB}) must also be transverse. This matches exactly our constraint $\nabla \cdot \mathbf{J}(\mathbf{r}) = 0$, which translates to $\mathbf{k}\cdot \overline{\mathbf{J}}(\mathbf{k}) = 0$ in $\mathbf{k}$-space, meaning that the current as well as the magnetization densities contributing to the helicity stored in a stationary system are transverse. Dividing by $|\mathbf{k}|$, then applying $\hat{\mathbf{k}}\times$ on both sides of Eq.~(\ref{eq:ME_k_curlB}) and using the vector identity $\mathbf{a}\times(\mathbf{b}\times\mathbf{c}) = \mathbf{b}(\mathbf{a}\cdot\mathbf{c}) - \mathbf{c}(\mathbf{a}\cdot\mathbf{b})$ gives us an expression for the magnetic field produced by a set of stationary sources $\overline{\mathbf{M}} (\mathbf{k})$ and $\overline{\mathbf{J}} (\mathbf{k})$
\begin{equation}
\label{eq:B_k}
    \overline{\mathbf{B}} (\mathbf{k}) = \mu_0 \overline{\mathbf{M}}_\perp (\mathbf{k})  + \mu _0 \text{i}\frac{\hat{\mathbf{k}}}{|\mathbf{k}|} \times\overline{\mathbf{J}} (\mathbf{k}) \, .
\end{equation}
The term $ \overline{\mathbf{M}}_\perp (\mathbf{k})$ denotes the transverse part of $\overline{\mathbf{M}} (\mathbf{k})$. We have written it explicitly in the above expression but will leave it out in the following because it is not relevant for the helicity, as all longitudinal parts vanish upon projection onto the transverse vectors $\hat{\mathbf{k}}\times\overline{\mathbf{M}} (\mathbf{k})$ and $\overline{\mathbf{J}} (\mathbf{k})$.
The helicity stored in a stationary system then becomes
\begin{widetext}
\begin{equation}
    \label{eq:helicity_magnetic_k}
    \begin{aligned}
            \braket{\Lambda} &=  \int _{\mathbb{R}^3} \frac{\mathrm{d}^3 k}{\hbar c \mu_0 |\mathbf{k}|} \overline{\mathbf{B}}^\dagger(\mathbf{k})\big( \text{i}\hbar\hat{\mathbf{k}}\times  \overline{\mathbf{B}}(\mathbf{k})  \big) \\
            &=  \int _{\mathbb{R}^3} \frac{\mathrm{d}^3 k}{\hbar c \mu_0 |\mathbf{k}|} \bigg(
            \mu_0\overline{\mathbf{M}}^\dagger (\mathbf{k})  + \mu_0\Big( \text{i} \frac{\hat{\mathbf{k}}}{|\mathbf{k}|} \times\overline{\mathbf{J}} (\mathbf{k}) \Big)^\dagger \bigg)
            \bigg( \mu_0  \text{i}\hbar \hat{\mathbf{k}} \times \overline{\mathbf{M}} (\mathbf{k}) + \frac{\mu_0\hbar}{|\mathbf{k}|}\overline{\mathbf{J}}(\mathbf{k})  \bigg) \\
            &= \int _{\mathbb{R}^3} \frac{\mathrm{d}^3 k}{ c |\mathbf{k}|} \mu_0 \bigg( \overline{\mathbf{M}}^\dagger (\mathbf{k}) \big( \text{i} \hat{\mathbf{k}} \times \overline{\mathbf{M}} (\mathbf{k})  \big) + \frac{1}{|\mathbf{k}|^2} \Big( \text{i} \hat{\mathbf{k}}\times\overline{\mathbf{J}} (\mathbf{k}) \Big) ^\dagger \,  \overline{\mathbf{J}} (\mathbf{k})  +  \frac{1}{|\mathbf{k}|} \big( \overline{\mathbf{M}}^\dagger (\mathbf{k})  \, \overline{\mathbf{J}} (\mathbf{k}) + \overline{\mathbf{J}}^\dagger (\mathbf{k}) \, \overline{\mathbf{M}} (\mathbf{k}) \big) \bigg)
    \end{aligned}
\end{equation}
\end{widetext}
where in the first line, the expression in Eq.~(\ref{eq:helicity_generalform_E_B}) is simplified using Eq.~(\ref{eq:ME_k_curlE}), and then in the second line Eqs.~(\ref{eq:ME_k_curlB}) and~(\ref{eq:B_k}) were utilized to arrive at the three distinct terms in the expression for the helicity in the last step. 
We have a pure magnetization density term (the first term; we will call it $\Lambda_{M}$), a pure current density term (the second term, $\Lambda_{J}$), and a mixed term (the third term, $\Lambda_{JM}$). 
The pure terms, or self-helicities, quantize how the magnetic field lines associated with the source in question wind and twist around each other. 
The mixed term, or mutual helicity, analogously quantifies the relative winding of the field lines arising from the two different sources. 
With the final expression in Eq.~(\ref{eq:helicity_magnetic_k}), we can calculate the helicity stored in a stationary system with given magnetization and current densities. 
However, if a method for acquiring the magnetization and current densities of a molecular system is used, which yields these quantities in real space, i.e., $\mathbf{M}(\mathbf{r})$ and $\mathbf{J}(\mathbf{r})$, we would first have to acquire their Fourier transforms $\overline{\mathbf{M}} (\mathbf{k})$ and $\overline{\mathbf{J}} (\mathbf{k})$. 
In such cases, it can prove to be rather difficult numerically to find the right $\mathbf{k}$-vector values, which capture all relevant properties of the magnetization and current densities.
Then, one would rather have a formula for the helicity content, which directly includes $\mathbf{M}(\mathbf{r})$ and $\mathbf{J}(\mathbf{r})$ instead of $\overline{\mathbf{M}} (\mathbf{k})$ and $\overline{\mathbf{J}} (\mathbf{k})$. 
To that end, we rewrite the last step of Eq.~(\ref{eq:helicity_magnetic_k}) as a real-space integral
\begin{widetext}
\begin{equation} \label{eq:helicity_real_space}
\begin{aligned}
    \braket{\Lambda} &= \Lambda_{M} + \Lambda_{J} + \Lambda_{JM}\\ &= \frac{\mu_0}{ 4 \pi c}\int _{\mathbb{R}^3} \mathrm{d}^3r \int _{\mathbb{R}^3} \mathrm{d}^3r' \,\frac{(\mathbf{r}-\mathbf{r}')
    \cdot \big(\mathbf{M} (\mathbf{r}) \times \mathbf{M} (\mathbf{r}')\big) }{|\mathbf{r}-\mathbf{r}'|^3} \\
    & \ \ \ + \frac{\mu_0}{ 16 \pi^2 c}\int _{\mathbb{R}^3} \mathrm{d}^3r \, \bigg( \int _{\mathbb{R}^3} \mathrm{d}^3r' \, \frac{\mathbf{J}(\mathbf{r}') \times (\mathbf{r}-\mathbf{r}')}{|\mathbf{r}-\mathbf{r}'|^3}  \bigg) \cdot \bigg( \int _{\mathbb{R}^3} \mathrm{d}^3r'' \frac{\mathbf{J}(\mathbf{r}'')}{|\mathbf{r}-\mathbf{r}''|}  \bigg) \\
    & \ \ \ + \frac{\mu_0}{ 2 \pi c}\int _{\mathbb{R}^3} \mathrm{d}^3r \int _{\mathbb{R}^3} \mathrm{d}^3r' \, \frac{\mathbf{M} (\mathbf{r})\cdot  \mathbf{J} (\mathbf{r}') }{|\mathbf{r}-\mathbf{r}'|} \, ,
    \end{aligned}
\end{equation}
\end{widetext}
where we again have the three distinct terms $\Lambda_{M}$, $\Lambda_{J}$ and $\Lambda_{JM}$ of the helicity, in that order. 
The mathematical details regarding the transformation of the integrals from Eq.~(\ref{eq:helicity_magnetic_k}) to Eq.~(\ref{eq:helicity_real_space}), as well as alternative forms of Eq.~(\ref{eq:helicity_real_space}) can be found in Appendix~\ref{app:helicity_k_to_r}. 
Since $\mathbf{M}(\mathbf{r})$ and $\mathbf{J}(\mathbf{r})$ are vectorial quantities bounded to the volume of the system, all integrals over ${\mathbb{R}^3}$ collapse to integrals over the relevant volume $V$. 
Equation~(\ref{eq:helicity_real_space}) equips us with the most straightforward way to calculate the helicity stored in a stationary molecule with a real-space magnetization or current density distribution, or both.

An important note, which also serves as verification of our helicity expressions, is in order. 
The helicity content of stationary systems with magnetization and current densities derived in this section, Eqs.~(\ref{eq:helicity_magnetic_k}) and~(\ref{eq:helicity_real_space}), corresponds to the magnetic helicity of these systems. 
As explained in Sec.~\ref{sec:introduction}, magnetic helicity is a quantity that characterizes the topological properties of magnetic field lines -- their spatial twistedness and knottedness -- and is related to the Hopf charge. 
It is given by the volume integral of the scalar product of the magnetic field $\mathbf{B}(\mathbf{r})$ with the magnetic vector potential $\mathbf{A}(\mathbf{r})$. 
If we take Maxwell's equations in the stationary limit, Eq.~(\ref{eq:Maxwell_stationary_r}), substitute $\mathbf{B}(\mathbf{r}) = \nabla \times\mathbf{A}(\mathbf{r})$, and choose the Coulomb gauge, where $\nabla \cdot \mathbf{A}(\mathbf{r}) = 0$, we get 
\begin{equation}
    \label{eq:laplace_A}
    \nabla ^2 \mathbf{A}(\mathbf{r}) = -\mu_0 \nabla \times \mathbf{M}(\mathbf{r}) - \mu_0 \mathbf{J}(\mathbf{r}) \, ,
\end{equation}
where $\nabla ^2$ is the Laplacian in $\mathbb{R}^3$. 
The above inhomogeneous differential equation for $\mathbf{A}(\mathbf{r})$ can be solved using the free-space Green's function of the Laplacian to yield \cite[Chapter~10]{arfkenMathematicalMethodsPhysicists2013}
\begin{equation}
\begin{aligned}
    \label{eq:A(r)}
    \mathbf{A}(\mathbf{r}) = & \frac{\mu_0}{4\pi} \int _{\mathbb{R}^3} \mathrm{d}^3r'\, \frac{\mathbf{J}(\mathbf{r'})}{|\mathbf{r}-\mathbf{r'}|} \\ &+  \frac{\mu_0}{4\pi} \int _{\mathbb{R}^3} \mathrm{d}^3r' \, \frac{\nabla' \times \mathbf{M}(\mathbf{r'})}{|\mathbf{r}-\mathbf{r'}|} \, ,
\end{aligned}
\end{equation}
where $\nabla '$ is the nabla operator with respect to the primed coordinates $\mathbf{r}'$. 
It can then be shown that 
\begin{equation}
    \label{eq:helicity_equivalence}
    \braket{\Lambda} = \frac{1}{\mu_0 c} \int _{\mathbb{R}^3} \mathrm{d}^3r \, \mathbf{B}(\mathbf{r})\cdot\mathbf{A}(\mathbf{r}) \, .
\end{equation}
That means that our helicity expressions, Eqs.~(\ref{eq:helicity_magnetic_k}) and~(\ref{eq:helicity_real_space}), are, in fact, equivalent to the magnetic helicity stored in a magnetic field arising from magnetization and current densities (the integral in Eq. (\ref{eq:helicity_equivalence})). 
The two differ only by a proportionality factor of ${\mu_0 c}={Z_0}$, where $Z_0$ is the vacuum impedance, which is a result of the helicity having slightly different definitions and, therefore, different units. 
We note that our definition for the helicity leads to it having units of angular momentum, or $\hbar$. 
We prove the equivalence of our helicity expressions and magnetic helicity in Appendix~\ref{app:magnetic_helicity}. 

\section{Numerical Results}
\label{sec:results}
Equipped with the formalism to calculate the helicity content of magnetization and current density configurations, we move on to its numerical application to stationary chiral molecules in this section. First, we summarize the numerical methods used to acquire the needed quantities.

\subsection{Numerical Methods}
\label{subsec:methods}
We use spin-current-dependent non-collinear density functional 
theory \cite{Van-Wullen:Spin.2002,holzerCurrentDensityFunctional2022,Pausch.Holzer:Linear.2022},
abbreviated as sDFT, as a basis to describe the electron density.
From the occupied spinors $\varphi_j$, we evaluate on each grid point $\mathbf{r}$
the electron density $\rho$, the spin magnetization density $\mathbf{m}$ with
Cartesian components $m_u$ ($u \in \{x,y,z\}$), and the (canonical)
paramagnetic current density $\mathbf{j}_{\text{p}}$ with components $j_{\text{p},\alpha}$ \cite{tellgrenDensityfunctionalTheoryInternal2018,sundholmCurrentDensityMolecular2021,holzerCurrentDensityFunctional2022}
\begin{align}
  \rho(\mathbf{r}) & = \sum_j \varphi_j^{\dagger} \varphi_j , \label{eq:rho} \\
  m_u(\mathbf{r}) & = \sum_j \varphi_j^{\dagger} \sigma_u \varphi_j , \label{eq:m} \\
  j_{\text{p},u}(\mathbf{r}) & = -\frac{\mathrm{i}}{2} \sum_j
     \left[ \varphi_j^{\dagger} (\nabla_{u} \varphi_j)
          - (\nabla_{u} \varphi_j)^{\dagger} \varphi_j \right] , \label{eq:jp}
\end{align}
where $\sigma_u$ are the Pauli spin matrices. All quantities are given in atomic units. 
If a static external magnetic field is turned on, neglecting Eq.~(\ref{eq:jp}) 
at the DFT level will lead to a strong gauge-variance of the system within the magnetic field. 
We note that only the physical current 
$\mathbf{j}_A(\mathbf{r})=\mathbf{j}_{\text{p}}(\mathbf{r})+\mathbf{j}_{\text{d}}(\mathbf{r})$ 
with $j_{\text{d}}(\mathbf{r}) =-\rho(\mathbf{r}) \mathbf A(\mathbf{r})$ is gauge-invariant by itself, 
while $\mathbf{j}_{\text{p}}$ is also required to make the kinetic energy density gauge invariant \cite{Dobson:Alternative.1993}. 
This is in addition to the London atomic orbitals used to ensure gauge-origin independence 
in finite-field Hartree–Fock and (current-)density functional calculations \cite{Tellgren.Soncini.ea:Nonperturbative.2008,Tellgren.Teale.ea:Non-perturbative.2014}. 
We use $\mathbf{j}_{A}(\mathbf{r})$ as our current density for the helicity calculations. 
We note that $\mathbf{j}_A(\mathbf{r})=\mathbf{j}_{\text{p}}(\mathbf{r})$ in the absence of an applied magnetic field.

Some limiting cases deserve a comment. 
Closed-shell molecules without an external field are time-reversal symmetric (Kramers-restricted), hence $\mathbf{m}(\mathbf{r}) = 0$ everywhere. 
Also, we have $\mathbf{j}_{\text{p}}(\mathbf{r}) = 0$ in this limit, whereas $\rho(\mathbf{r})$ and its symmetric derivatives are generally non-zero. 
Likewise, if spin--orbit coupling (SOC) is negligible, which is the case for light elements since SOC scales approximately with $Z^4$, and no external magnetic field is present, the orbitals can be chosen real. 
The imaginary parts in Eqs.~(\ref{eq:jp}) then cancel and $\mathbf{j}_{\text{p}}(\mathbf{r}) = 0$. 
In an external magnetic field $\mathbf{B}_\text{ext}$, however, the Zeeman and orbital terms make the Hamiltonian complex, so a finite current density arises even without SOC \cite{Pausch.Holzer:Linear.2022}.

The acquired real space magnetization and current densities are brought from atomic units to SI units as
\begin{equation}
\label{eq:au_to_SI_conversion}
    \mathbf{M}(\mathbf{r})^{\text{SI}} = \frac{-g_e\mu_B}{2a_0 ^3}\mathbf{m}(\mathbf{r})^{\text{au}} \, , \ \mathbf{J}(\mathbf{r})^{\text{SI}} = \frac{-
    e E_{\text{h}}}{\hbar a_0^2} \mathbf{j}_A(\mathbf{r})^{\text{au}} \, ,
\end{equation}
where $g_e$ is the electron $g$-factor, $\mu_B$ the Bohr magneton, $a_0$ the Bohr radius, $e$ the elementary electric charge, and $E_{\text{h}}$ the Hartree energy. 
The quantities on the right-hand sides of Eq.~(\ref{eq:au_to_SI_conversion}) are used to calculate the helicity content of the different molecular configurations using Eq.~(\ref{eq:helicity_real_space}). 
The quantities are evaluated on an irregular grid with an effective volume associated with each point and, hence, any volume integral simplifies to a weighted sum. 

The singular points in the helicity integrals are treated using the self-term correction method \cite{livesayElectromagneticFieldsInduced1974}. 
Hence, we first carry out the discretized version of the integral, excluding the points where $\mathbf{r}=\mathbf{r}'$, and then add back the local self-term correction corresponding to the $\mathbf{r}=\mathbf{r}'$ contribution. 
For example, the self-term contribution for the inner integral of $\Lambda_{JM}$ becomes
\begin{equation}
    \begin{aligned}
        &\int _{B_\eta(\mathbf{r})} \mathrm{d}^3r' \, \frac{\mathbf{M} (\mathbf{r})\cdot  \mathbf{J} (\mathbf{r}') }{|\mathbf{r}-\mathbf{r}'|} \\&\approx \mathbf{M} (\mathbf{r})\cdot  \mathbf{J} (\mathbf{r})  \int _{B_\eta(\mathbf{r})} \mathrm{d}^3r' \, \frac{1}{|\mathbf{r}-\mathbf{r}'|}\\
        &= \mathbf{M} (\mathbf{r})\cdot  \mathbf{J} (\mathbf{r}) \, 2\pi \eta^2 \, ,
    \end{aligned}
\end{equation}
where in the first line we have an integration over a ball of radius $\eta$ centered around the point $\mathbf{r}$. 
The assumption here is that we can approximate the irregular cell at $\mathbf{r}=\mathbf{r}'$ as a sphere of the same volume $V(\mathbf{r})=\frac{4}{3}\pi \eta^3$. 
Then, in the second line, we use the approximation $\mathbf{J}(\mathbf{r}') \approx \mathbf{J}(\mathbf{r})$ over the sphere, which lets us solve the remaining integral analytically, yielding $2\pi \eta^2$ with $\eta = \sqrt[3]{\frac{3}{4\pi}V(\mathbf{r})}$. 
For $\Lambda_M$, we do not have such a self-term correction, as $\mathbf{M}(\mathbf{r})\times\mathbf{M}(\mathbf{r}) = 0$. 
Finally, $\Lambda_J$ would have two such corrections as it contains two integrals with singular kernels. 
However, the self-term correction for the $\mathrm{d}^3 r'$ integral vanishes since $\int \mathrm{d}^3s \, \frac{\mathbf{s}}{|\mathbf{s}|^3} = 0$. 
The correction for the $\mathrm{d}^3 r''$ integral is analogous to the example above. 

The radial convergence of the helicity integrand is then verified using an analysis of the cumulative sum starting from the origin and going outwards. The good convergence behavior guarantees that a sufficiently large spatial domain around the molecule is considered and that the total helicity in fact comes from the molecule and is not a result of numerical noise far away from it. 

\subsection{BINOL}
\label{subsec:binol}
\begin{figure*} 
\centering
\includegraphics[width=1.0\textwidth]{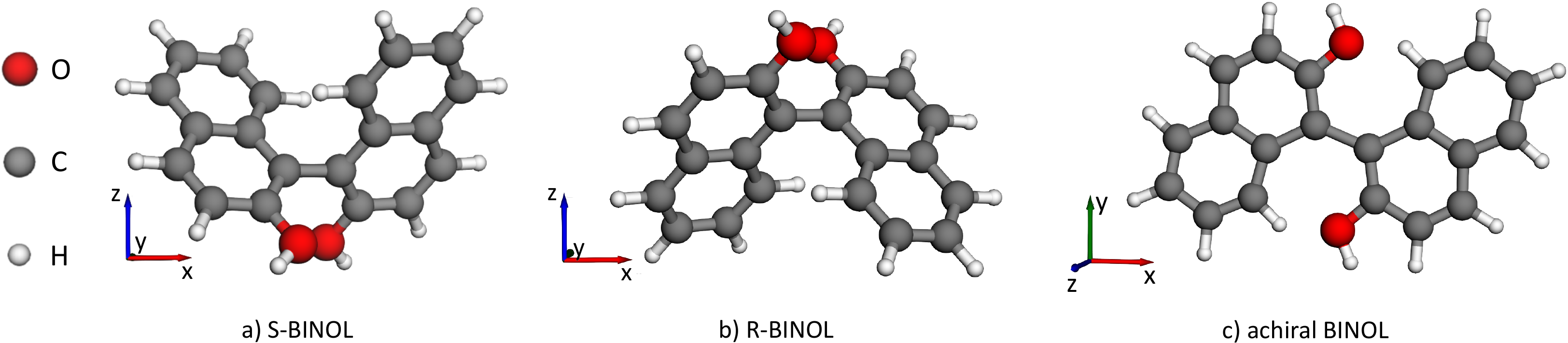}
\caption{Visualization of three BINOL variants: a) S-BINOL, b) R-BINOL, and c) achiral BINOL. The legend for the atoms is given on the left and is valid for all figures.}
\label{fig:BINOL_visualization}
\end{figure*}
The first molecule we study is C$_{20}$H$_{14}$O$_2$ (BINOL) in both of its chiral variants (S-BINOL and R-BINOL) as well as an achiral configuration obtained by enforcing strict C$_i$ point group symmetry, which includes spatial inversion as a molecular symmetry, which can be seen in Fig.~\ref{fig:BINOL_visualization}. 
BINOL is diamagnetic and has negligible spin-orbit coupling due to being composed only of light atoms. 
Hence, it has neither $\mathbf{M}(\mathbf{r})$ nor $\mathbf{J}(\mathbf{r})$ on its own. 
We place it in a static homogeneous external magnetic field $\mathbf{B}_\text{ext}$ to induce stationary currents, similar to prior works, where magnetically induced molecular current densities have previously been investigated \cite{Soncini.Teale.ea:Maps.2008}. 
The magnetic field in that case has an effect orders of magnitude larger than the spin-orbit coupling. 
We align the external magnetic field with the $z$-axis -- the axis about which BINOL has a $C_2$ rotational symmetry. 
Then, the combined system still has a $C_2$ rotational symmetry about the $z$-axis. 
The pure current term of the helicity ($\Lambda_J$ in Eq.~(\ref{eq:helicity_real_space})) is the only relevant contribution for BINOL in an external magnetic field. 
We vary the field strength, which affects the currents, and study the change of the helicity stored in BINOL. 
We keep to field strengths for which BINOL remains diamagnetic, that is, it is in a closed-shell singlet state. For magnetic field strengths of \SI{10000}{\tesla} and above, BINOL is paramagnetic -- it is in an open-shell triplet state. 
This is due to the Zeeman shift of the triplet energy levels, which drop linearly with the magnetic field strength. 
At some magnetic field strength value, the triplet energy drops below the singlet energy, making the former the new ground state. 
Our triplet simulations yield partially unphysical results from a symmetry standpoint, and we will, hence, not explore the open-shell state of BINOL in this work.

\begin{figure}
    \centering
    \includegraphics[width=\linewidth]{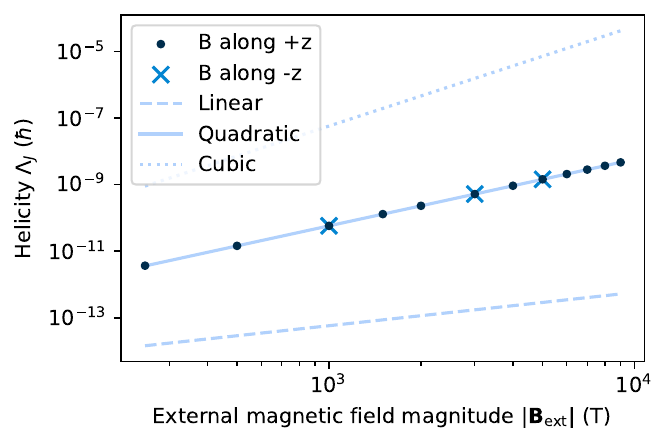}
    \caption{Change of the helicity stored in S-BINOL with respect to the magnitude of the external magnetic field. The dots represent the helicity values of S-BINOL for different field magnitudes when the magnetic field points along $\hat{z}$, while the crosses correspond to those values when the field points along $-\hat{z}$. As seen by the comparison of the three lines plotted for reference in the figure, representing a linear (dashed line), quadratic (solid line), and cubic function (dotted line), we observe a quadratic dependence of $\Lambda_J$ on $|\mathbf{B}_\text{ext}|$.}
    \label{fig:hel_wrt_B}
\end{figure}

Figure~\ref{fig:hel_wrt_B} shows that the helicity stored in S-BINOL grows quadratically with the strength of the external magnetic field, going from $\SI{3.673e-12}{\hbar}$ at $|\mathbf{B}_\text{ext}| = \SI{250}{\tesla}$ to $\SI{4.672e-9}{\hbar}$ at $|\mathbf{B}_\text{ext}| = \SI{9000}{\tesla}$. 
We get practically identical helicity values for magnetic fields of the same magnitude but opposite direction. 
The square dependence is recognized through the excellent agreement of the data with the quadratic function plotted for reference in Fig.~\ref{fig:hel_wrt_B}; a power law fit yields a power of $1.994$. 
In this range of external magnetic field values, we expect a linear behavior of the current density with respect to $\mathbf{B}_\text{ext}$. 
Without a field, the (closed-shell) molecule carries no current, 
and since the current density is odd under time reversal, 
$\mathbf{J}(-\mathbf{B_\text{ext}}) = -\mathbf{J}(\mathbf{B}_\text{ext})$, only odd powers of
$\mathbf{B}_\text{ext}$ can occur in a series expansion of $\mathbf{J}(\mathbf{B}_\text{ext})$ around $\mathbf{B}_\text{ext}=\mathbf{0}$. The leading term is thus linear, as the used magnetic field strength measured in atomic units is weak compared to the Coulomb force acting on the electrons in the molecule. 
In turn, the current density enters the helicity term $\Lambda_J$ in Eq.~(\ref{eq:helicity_real_space}) quadratically, accounting for the overall observed dependence of the helicity on the external magnetic field, including the data points, for which the magnetic field points in the opposite direction. 
The data for R-BINOL yields, within noise levels, the same helicity values but with a minus sign, perfectly representing the pseudoscalar nature of the helicity, which changes sign under spatial inversion. 
The achiral configuration of BINOL is used as a test case, as we expect it to not contain any helicity due to its inversion symmetry. 
The data entirely agrees with this reasoning by yielding zero helicity for all external magnetic field strengths, up to numerical noise.

\begin{figure*}
    \centering
    \includegraphics[width=1.0\linewidth]{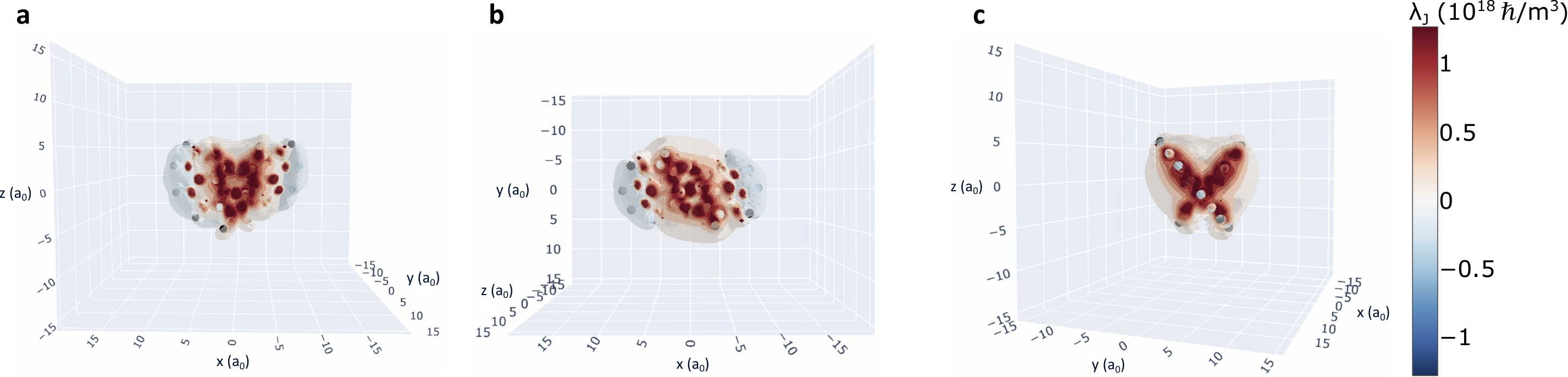}
    \caption{Spatial distribution of the helicity integrand denoted by $\lambda_J$ of S-BINOL in a static homogeneous external magnetic field $|\mathbf{B}_\text{ext}| = \SI{1000}{\tesla}$ along $+\hat{z}$ from three different perspectives as denoted by the axis labels, which mark the orientation of the coordinate system. The coordinates are given in units of the Bohr radius $a_0$. The color bar on the right applies to all three images.}
    \label{fig:helicity_integrand}
\end{figure*}

In Fig.~\ref{fig:helicity_integrand}, we show the spatial distribution of the helicity integrand for S-BINOL in $|\mathbf{B}_\text{ext}|=\SI{1000}{\tesla}$. 
We note that the shown pattern does not change for a different magnetic field strength, the values simply scale quadratically with $|\mathbf{B}_\text{ext}|$, and for R-BINOL, the pattern is mirrored and switches sign. 
The helicity integrand for the achiral BINOL version consists of numerical zeros. 

In Fig.~\ref{fig:helicity_integrand}, we see predominantly regions of positive helicity, with two slightly negative regions on the outer sides of the molecule. 
The major contribution to the helicity of BINOL comes from the central parts of the two flaps that make it up. 
The relative twist between these flaps is what makes BINOL chiral, and it hence seems natural for the currents, which follow the geometry, and the associated magnetic field to store the most helicity in the inner region of the flaps. 
Furthermore, the spatial distribution of the helicity integrand respects the $C_2$ rotational symmetry of the system around the $z$-axis, which is an important consistency check. 

Overall, BINOL in an external static homogeneous magnetic field exhibits helicity values ranging in orders of magnitude from $10^{-12}$ to $10^{-9}$ for the external field strengths we have explored here. 
In the following, we study a more complicated molecular system, which has both a current density as well as a non-zero magnetization distribution.

\subsection{Chiral Single-Molecule Magnet}
\label{subsec:europium_magnet} 
\begin{figure}
    \centering
    \includegraphics[width=\linewidth]{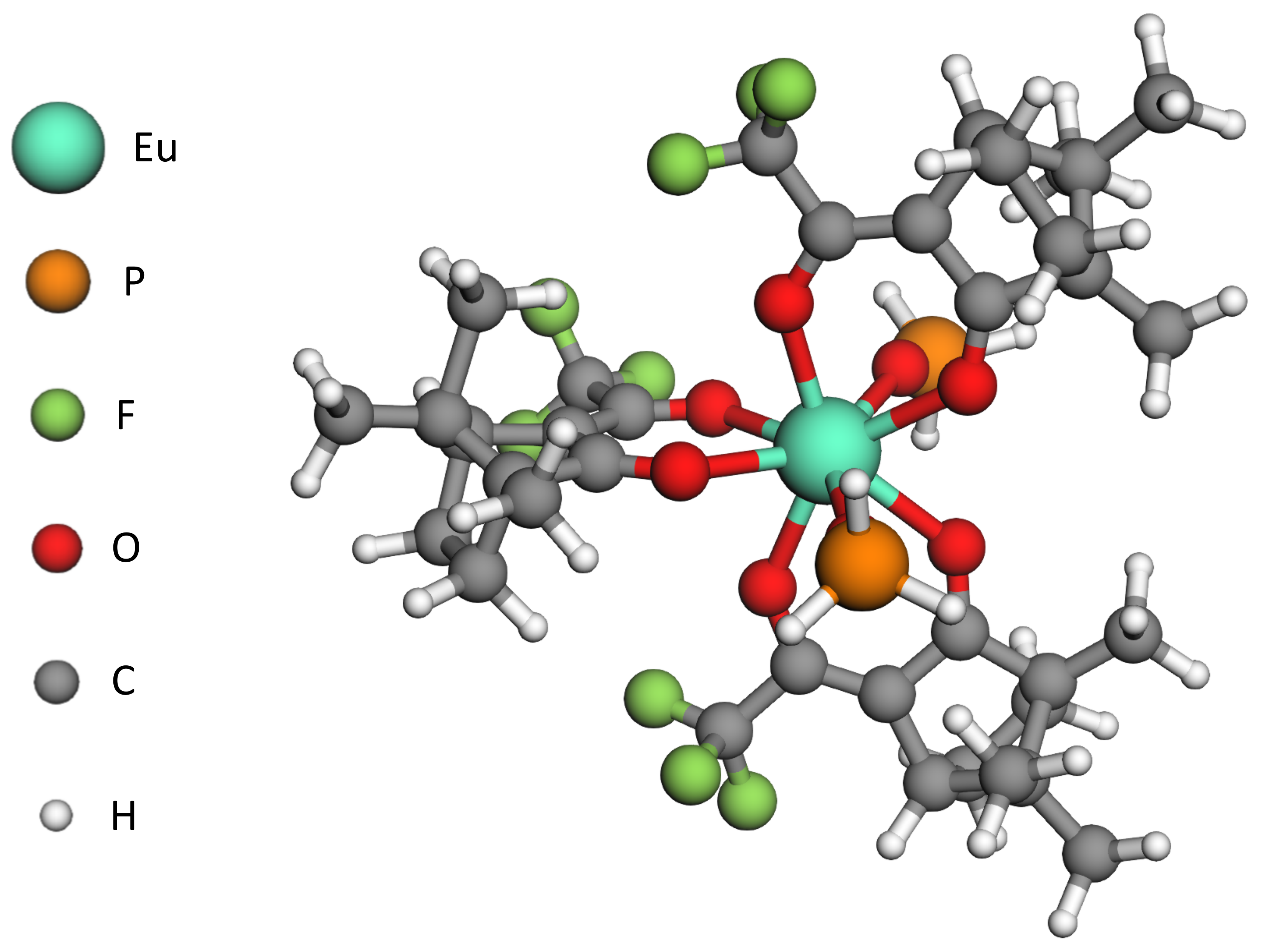}
    \caption{Visualization of the chiral molecular magnet with a Europium core [Eu(tfc)$_3$(OPH$_3$)$_2$]. The legend for the different atoms is given on the left.}
    \label{fig:Eu_magnet}
\end{figure}

We now turn to a different molecule, which has both currents due to spin-orbit coupling and magnetization coming from electron spin. 
As a model system, we use the phosphine oxide trifluoro camphor derivative [Eu(tfc)$_3$(OPH$_3$)$_2$], 
as a slightly simplified model system for a known strong circular polarized light emitter \cite{Harada.Nakano.ea:Circularly.2009A}. 
It is a chiral open-shell molecule with a Europium core, as visualized in Fig.~\ref{fig:Eu_magnet}, and its size is about $\SI{1.4}{\nano\metre}$. 
In typical Lanthanide fashion, this metal center allows it to function as a single-molecule magnet, that is, have a non-zero magnetization density. 
As Europium is heavy, the molecule experiences considerable spin-orbit coupling, which leads to emerging stationary currents. 
In this case, both $\mathbf{M}(\mathbf{r})$ and $\mathbf{J}(\mathbf{r})$, or equivalently, the magnetic fields associated with them, can store helicity. 
Expressed in terms of the magnetization and current density, the total helicity is comprised of all three terms in Eq.~(\ref{eq:helicity_real_space}) as given in Tab.~\ref{tab:eu_helicity}. 
The biggest contribution to the total helicity of $\SI{1.427e-09}{\hbar}$ is the mixed current--magnetization term $\Lambda_{JM}$ with $\SI{1.712e-09}{\hbar}$. 
Through an analysis of the spatial distribution of the $\Lambda_{JM}$ integrand, we find that $99\%$ of the final value is already reached within a radius of $\SI{0.1}{\nano\metre}$ around the Europium core. 
The source of the magnetization density is the unpaired electrons in the $4f$ orbitals of the molecule, which generally feature a very low radial overlap with the surrounding ligands, therefore remaining mostly in an atomic state. 
Ligands, such as the ones in [Eu(tfc)$_3$(OPH$_3$)$_2$], that strongly and non-uniformly deform the electron density are therefore required to distort the $4f$ orbitals. 
Once the local symmetry of the Europium $4f$ shells is broken, the extensive coupling of spin and orbit in the $4f$ shell leads to large current-density driven effects. 
Indeed, we observe that the helicity of this molecule is high relative to that of BINOL, and approximately equal to the helicity stored in BINOL at an external magnetic field strength of $\SI{5000}{\tesla}$. 
Furthermore, it has both quantities that can store helicity -- magnetization and current density -- therefore getting all three helicity contributions. 
The two smaller contributions to the total helicity almost cancel each other out, both being of the order of $10^{-10}$ and having opposite signs. 
However small, they still slightly impact the total helicity, bringing it down, as they are both negative. 
The negative sign of the current term is not a contraction to the positive sign of the pure magnetization and mixed terms. 
It only means that the associated magnetic field lines of the current term wind around themselves in the opposite sense relative to those of the magnetization winding around themselves and to those of the magnetization and current winding around each other.

\begin{table}
    \centering
    \begin{tabular}{l|c|S[table-format=-2.3e-2]}
        Term & Symbol & {Value (in $\hbar$)} \\
         \hline
         Magnetization \ & $\Lambda_{M}$ & 7.148e-11\\
         Current \ & $\Lambda_{J}$ & -3.564e-10\\
         Mixed \ & $\Lambda_{JM}$ & 1.712e-09 \\
         \hline
         Total \ & $\Lambda$ & 1.427e-09 \\
    \end{tabular}
    \caption{Helicity stored in the chiral single-molecule magnet [Eu(tfc)$_3$(OPH$_3$)$_2$], broken down into the individual contributing terms.}
    \label{tab:eu_helicity}
\end{table}

\section{Conclusions}
\label{sec:conclusions}
In this article, we study the helicity stored in stationary chiral molecular systems. 
Such systems can have helicity coming from two different vector fields native to molecules -- magnetization density and current density. 
The former is related to electron spin, while the latter is naturally present in molecules with non-negligible spin-orbit coupling, but it can also be generated via an external magnetic field in case the spin-orbit coupling is not sufficiently large. 
We expand the theoretical formalism presented in previous works \cite{fernandez-corbatonTotalHelicityElectromagnetic2021,fernandez-corbatonScalarProductComputing2023} and arrive at formulas that can be readily used to calculate the helicity stored in a system given its magnetization and current densities directly either in $\mathbf{k}$-space or real space. 
We show that our formulas correspond to a topological property known as magnetic helicity, thereby validating them. 
We use non-collinear spin density functional theory with the software TURBOMOLE to simulate the systems we study and get the magnetization and current density. 

We find that when BINOL is placed in an external homogeneous magnetic field, the induced stationary currents scale linearly in the diamagnetic regime. 
The helicity stored in BINOL then scales quadratically with the strength of the applied magnetic field, ranging from $\SI{3.67e-12}{\hbar}$ at $\SI{250}{\tesla}$ to $\SI{4.675e-09}{\hbar}$ at $\SI{9000}{\tesla}$. 
Furthermore, we show that the spatial distribution of the helicity integrand respects the $C_2$ rotational symmetry of the whole system. We then study a larger and more complicated open-shell molecule with a metal Europium core, which considers all contributions to the helicity. 
The total helicity of the Europium-based molecule amounts to $\SI{1.427e-09}{\hbar}$, which is about the same as the helicity value found for BINOL in an external magnetic field of $\SI{5000}{\tesla}$. 
We find that the largest and most important contribution is that from the mixed current--magnetization term, and it is mostly coming from the first $\SI{0.1}{\nano\metre}$ around the core. 
This points to an interplay between the unpaired electrons close to the core and the chiral ligands attached to the Europium atom in the center. 
The combination results in the lack of inversion symmetry of the molecule being felt most strongly around the central Europium atom, and in a large helicity being stored in that region. One may compare the helicity in the Europium-based molecule, equal to $\SI{1.427e-09}{\hbar}$, to the helicity in a Hopfion hosted inside a FeGe magnetic disk of $\SI{70}{\nano\meter}$ in height and $\SI{210}{\nano\meter}$ in diameter, whose magnitude was computed in \cite{fernandez-corbatonScalarProductComputing2023} to be equal to $\SI{129.1}{\hbar}$.

We conclude the article with an outlook regarding the importance that the helicity stored in chiral molecules computed here might have for chiral light--matter interactions. 
It is known that chiral molecules interact differently with light of electromagnetic helicity (different polarization handedness) and that a free field typically has a different helicity content before and after interaction with matter. 
As shown in this article, the fundamental mathematical formulation is the same for the helicity of free fields (electromagnetic helicity) and the helicity of stationary molecules (magnetic helicity). 
The electromagnetic helicity is merely a sum of a magnetic and an electric helicity term, and for a stationary molecule, the electric helicity is identically zero. 
It has also been shown that magnetic helicity can also be formulated using the same operator and the same scalar product as that for electromagnetic helicity \cite{fernandez-corbatonScalarProductComputing2023,grossNormInvarianceMassZero1964, bialynicki-birulaPhotonWaveFunction1996}. 
The two could, hence, be considered as two forms of the same fundamental quantity -- helicity \cite{fernandez-corbatonTotalHelicityElectromagnetic2021}. 
The possibility naturally raises the question of whether there is, or can be, helicity transfer between light and matter upon interaction. 
To find an answer, one needs to exit the stationary regime and simulate such an interaction, resolving the evolution of the electron spin and the currents in the molecule. 
Such explorations are left as a possible continuation of this work.

\section*{Acknowledgments}
I.I. thanks Jan David ``JD'' Fischbach for helpful discussions regarding this work. 
I.I. and L.R. acknowledge support by the Karlsruhe School of Optics \& Photonics (KSOP) and the Ministry of Science, Research and Arts of Baden-Württemberg as part of the sustainability financing of the projects of the Excellence Initiative II. 
I.I., L.R., C.R., and I.F.-C. acknowledge support by the Deutsche Forschungsgemeinschaft (DFG, German Research Foundation) -- Project-ID 258734477 -- SFB 1173. 

Claude Opus 5.5 was used in this work to a limited extent to independently cross-check analytical derivations and numerical results that had first been obtained by the authors, and to assist in debugging Python code. 
All derivations and code were developed by the authors, and all AI output was verified by the authors.
No AI tools were used for writing or editing the manuscript, reviewing the literature, or interpreting the results.

\appendix

\section{Helicity Integrals in Real Space}

\label{app:helicity_k_to_r}

We start by listing some useful formulas for transforming integrals between real and $\mathbf{k}$-space, which hold for our symmetric Fourier transform convention:
\begin{widetext}
\begin{subequations}
    \begin{align}
        &\int \mathrm{d}^3 k \, \overline{\mathbf{a}}^*(\mathbf{k}) \cdot \overline{\mathbf{b}}(\mathbf{k}) = \int \mathrm{d}^3 r \, {\mathbf{a}}^*(\mathbf{r}) \cdot \overline{\mathbf{b}}(\mathbf{r}) \ , \label{app_eq:parseval} \\
        &\int \mathrm{d}^3 k \, \overline{f}(\mathbf{k}) \, \overline{\mathbf{a}}^*(\mathbf{k}) \cdot \overline{\mathbf{b}}(\mathbf{k}) = \frac{1}{\sqrt{(2\pi)^3}} \int \mathrm{d}^3 r \, \int \mathrm{d}^3 r' \, f(\mathbf{r} - \mathbf{r}') \, {\mathbf{a}}^*(\mathbf{r}) \cdot \overline{\mathbf{b}}(\mathbf{r}') \ ,\label{app_eq:generalized_parseval} \\
        &\int \mathrm{d}^3 k \, \overline{\mathbf{c}}(\mathbf{k}) \cdot (\overline{\mathbf{a}}^*(\mathbf{k}) \times \overline{\mathbf{b}}(\mathbf{k})) = \frac{1}{\sqrt{(2\pi)^3}} \int \mathrm{d}^3 r \, \int \mathrm{d}^3 r' \, \mathbf{c}(\mathbf{r} - \mathbf{r}') \cdot ({\mathbf{a}}^*(\mathbf{r}) \times \overline{\mathbf{b}}(\mathbf{r}')) \ , \label{app_eq:generalized_parseval_cross}
    \end{align}
\end{subequations}    
\end{widetext}
where the $\mathbf{k}$-space representation of some scalar real-space function $f(\mathbf{r})$ is denoted by $\overline{f}(\mathbf{k})$ and equivalently for some vector-valued functions $\mathbf{a}$, $\mathbf{b}$ and $\mathbf{c}$. 
Equation~(\ref{app_eq:parseval}) is the Plancherel formula, and Eqs.~(\ref{app_eq:generalized_parseval}) and~(\ref{app_eq:generalized_parseval_cross}) are generalizations thereof. 
The proof of the above formulas is straightforward -- one substitutes the definitions of the Fourier transforms of the $\mathbf{k}$-space quantities and then removes the integral over $\mathrm{d}^3 k$ either by solving it directly or by identifying it as part of an inverse Fourier transform. 
We will make use of the following Fourier transform pairs
\begin{subequations}
\label{app_eq:ft_identities}
    \begin{align}
        \text{i}\mathbf{k}\overline{{f}}(\mathbf{k}) &\leftrightarrow \nabla {{f}}(\mathbf{r}) \, , \label{app_eq:FT_grad_f}\\
        \text{i}\mathbf{k}\times \overline{\mathbf{a}}(\mathbf{k}) &\leftrightarrow \nabla \times {\mathbf{a}}(\mathbf{r}) \, , \label{app_eq:FT_curl_A}\\
        \overline{{f}}(\mathbf{k})\overline{\mathbf{a}}(\mathbf{k}) &\leftrightarrow \frac{1}{\sqrt{(2\pi)^3}} \int _{\mathbb{R}^3} \mathrm{d}^3 r' \, f(\mathbf{r}-\mathbf{r}') \mathbf{a}(\mathbf{r}') \, , \label{app_eq:FT_convolution}\\
        \frac{1}{|\mathbf{k}|^2} &\leftrightarrow  \sqrt{\frac{\pi}{2}}\frac{1}{|\mathbf{r}|} \, , \label{app_eq:FT_1/k2}
    \end{align}
\end{subequations}
where the third pair is just the convolution theorem. 
We will go through the integral transformations one term at a time. For the pure magnetization term, $\Lambda_M$, we have
\begin{widetext}
\begin{equation}
\label{app_eq:lambdaM_McurlM_k_to_r}
    \begin{aligned}
        \Lambda_M &=  \int _{\mathbb{R}^3} \frac{\mathrm{d}^3 k}{ c |\mathbf{k}|} \mu_0 \bigg( \overline{\mathbf{M}}^\dagger (\mathbf{k}) \big( \text{i} \hat{\mathbf{k}} \times \overline{\mathbf{M}} (\mathbf{k})  \big) \bigg) \\
        &= \frac{\mu _0}{c} \int _{\mathbb{R}^3} \mathrm{d}^3 k \, \frac{1}{ |\mathbf{k}|^2} \bigg( \overline{\mathbf{M}}^* (\mathbf{k}) \cdot \big( \text{i}{\mathbf{k}} \times \overline{\mathbf{M}} (\mathbf{k})  \big) \bigg) \\
        &=  \frac{\mu _0}{c} \frac{1}{\sqrt{(2\pi)^3}}\int _{\mathbb{R}^3} \mathrm{d}^3 r \int _{\mathbb{R}^3} \mathrm{d}^3 r' \,\sqrt{\frac{\pi}{2}}\frac{1}{|\mathbf{r}-\mathbf{r}'|} {\mathbf{M}} ^* (\mathbf{r}) \cdot  \big(\nabla ' \times {\mathbf{M}}(\mathbf{r}') \big) \\
        &= \frac{\mu_0}{ 4 \pi c}\int _{\mathbb{R}^3} \mathrm{d}^3r \int _{\mathbb{R}^3} \mathrm{d}^3r' \,\frac{\mathbf{M} (\mathbf{r}')\cdot (\nabla \times \mathbf{M} (\mathbf{r})) }{|\mathbf{r}-\mathbf{r}'|} \ .
    \end{aligned}
\end{equation}
\end{widetext}
In the above equation, the steps go as follows. 
The form of $\Lambda_M$, taken from Eq.~(\ref{eq:helicity_magnetic_k}), is rewritten such as to visually separate the expression into a scalar function, $\overline{f}(\mathbf{k}) = \frac{1}{ |\mathbf{k}|^2}$, and the scalar product of two vector functions, $\overline{\mathbf{a}}^*(\mathbf{k}) =\overline{\mathbf{M}}^*(\mathbf{k})  $ and $\overline{\mathbf{b}}(\mathbf{k}) = \text{i}{\mathbf{k}} \times \overline{\mathbf{M}} (\mathbf{k})$. 
This separation allows us to use Eq.~(\ref{app_eq:generalized_parseval}) to rewrite the integral from $\mathbf{k}$-space to real space, utilizing the Fourier transform identities in Eqs.~(\ref{app_eq:FT_curl_A}) and~(\ref{app_eq:FT_1/k2}). 
Using the fact that $\mathbf{M}(\mathbf{r})$ is real and a variable switch $\mathbf{r}\leftrightarrow \mathbf{r}'$ for the sake of clearer notation leads us to the last line in the above equation. 
In some cases, especially when the data is coming from a simulation on an irregular grid, the curl of the magnetization might not be easy to obtain. 

Alternatively, we can arrive at an equivalent formula, which does not rely on the curl of the magnetization:
\begin{widetext}
\begin{equation}
\label{app_eq:lambdaM_MM_k_to_r}
    \begin{aligned}
        \Lambda_M &= \frac{\mu _0}{c} \int _{\mathbb{R}^3} \mathrm{d}^3 k \, \frac{\text{i}{\mathbf{k}} }{ |\mathbf{k}|^2} \cdot \big( \overline{\mathbf{M}} (\mathbf{k}) \times \overline{\mathbf{M}}^* (\mathbf{k})  \big) \\
        &=  \frac{\mu _0}{c} \frac{1}{\sqrt{(2\pi)^3}}\int _{\mathbb{R}^3} \mathrm{d}^3 r \int _{\mathbb{R}^3} \mathrm{d}^3 r' \,\Bigg(\sqrt{\frac{\pi}{2}}\nabla_{\mathbf{r}-\mathbf{r}'}\frac{1}{|\mathbf{r}-\mathbf{r}'|}\Bigg) \cdot \big( {\mathbf{M}} ^* (\mathbf{r}) \times {\mathbf{M}} (\mathbf{r}')  \big) \\
        &= \frac{\mu_0}{ 4 \pi c}\int _{\mathbb{R}^3} \mathrm{d}^3r \int _{\mathbb{R}^3} \mathrm{d}^3r' \,\frac{(\mathbf{r}-\mathbf{r}')
    \cdot \big(\mathbf{M} (\mathbf{r}) \times \mathbf{M} (\mathbf{r}')\big) }{|\mathbf{r}-\mathbf{r}'|^3} \ .
    \end{aligned}
\end{equation}
\end{widetext}
The steps taken above are similar to the ones in Eq.~(\ref{app_eq:lambdaM_McurlM_k_to_r}) after a transformation of the integrand using the vector identity $\mathbf{a}\cdot (\mathbf{b}\times \mathbf{c}) = \mathbf{b} \cdot (\mathbf{c}\times \mathbf{a})$ in the first line. 
There, we also visually separate the three functions, where here all three are vectorial -- $\overline{\mathbf{c}}(\mathbf{k}) = \frac{\text{i}{\mathbf{k}} }{ |\mathbf{k}|^2}$, $\overline{\mathbf{a}}^*(\mathbf{k}) = \overline{\mathbf{M}} (\mathbf{k})$ and $\overline{\mathbf{b}}(\mathbf{k}) = \overline{\mathbf{M}} ^* (\mathbf{k})$ -- allowing us to directly apply Eq.~(\ref{app_eq:generalized_parseval_cross}) with the help of Eqs.~(\ref{app_eq:FT_grad_f}) and~(\ref{app_eq:FT_1/k2}). 
Calculating the resulting gradient and again using the reality of $\mathbf{M}(\mathbf{r})$ leads us to the last line, which is also the form used for the numerical results in this work. 

The pure current density term of the helicity can be transformed to a real-space integral in an analogous way:
\begin{widetext}
\begin{equation}
\label{app_eq:lambdaJ_k_to_r}
    \begin{aligned}
        \Lambda_J &=  \int _{\mathbb{R}^3} \frac{\mathrm{d}^3 k}{ c |\mathbf{k}|^3} \mu_0  \Big( \text{i} \hat{\mathbf{k}}\times\overline{\mathbf{J}} (\mathbf{k}) \Big) ^\dagger \,  \overline{\mathbf{J}} (\mathbf{k}) \\
        &= \frac{\mu_0}{c}\int _{\mathbb{R}^3} \mathrm{d}^3 k\,  \Bigg( \text{i}{\mathbf{k}}\times\frac{\overline{\mathbf{J}} (\mathbf{k}) }{|\mathbf{k}|^2}\Bigg) ^* \cdot  \frac{\overline{\mathbf{J}} (\mathbf{k}) }{|\mathbf{k}|^2} \\
        &= \frac{\mu_0}{ c}\int _{\mathbb{R}^3} \mathrm{d}^3r \, \bigg( \nabla \times \frac{1}{4\pi} \int _{\mathbb{R}^3} \mathrm{d}^3r' \, \frac{\mathbf{J}(\mathbf{r}')}{|\mathbf{r}-\mathbf{r}'|}  \bigg)^* \cdot \bigg( \frac{1}{4\pi} \int _{\mathbb{R}^3} \mathrm{d}^3r'' \frac{\mathbf{J}(\mathbf{r}'')}{|\mathbf{r}-\mathbf{r}''|}  \bigg) \\
        &=  \frac{\mu_0}{ 16 \pi^2 c}\int _{\mathbb{R}^3} \mathrm{d}^3r \, \bigg( \int _{\mathbb{R}^3} \mathrm{d}^3r' \, \frac{\mathbf{J}(\mathbf{r}') \times (\mathbf{r}-\mathbf{r}')}{|\mathbf{r}-\mathbf{r}'|^3}  \bigg) \cdot \bigg( \int _{\mathbb{R}^3} \mathrm{d}^3r'' \frac{\mathbf{J}(\mathbf{r}'')}{|\mathbf{r}-\mathbf{r}''|}  \bigg) \, .
    \end{aligned}
\end{equation}
\end{widetext}
In the above equation, we have $\Lambda_J$ as it enters in Eq.~(\ref{eq:helicity_magnetic_k}) in the first line, then appropriately regrouped in the next one.
In that form, we can use Eq.~(\ref{app_eq:parseval}) to rewrite it as a real space integral in the following line, using also Eqs.~(\ref{app_eq:FT_curl_A}) and ~(\ref{app_eq:FT_convolution}). 
Applying the curl and using that $\mathbf{J}(\mathbf{r})$ is real, we arrive at the last step of Eq.~(\ref{app_eq:lambdaJ_k_to_r}), which is the form we use in the main article. 

Finally, for the mixed term, we have
\begin{widetext}
\begin{equation}
\label{app_eq:lambdaJM_k_to_r}
    \begin{aligned}
        \Lambda_{JM} &= \int _{\mathbb{R}^3} \frac{\mathrm{d}^3 k}{ c |\mathbf{k}|^2} \mu_0  \big( \overline{\mathbf{M}}^\dagger (\mathbf{k})  \, \overline{\mathbf{J}} (\mathbf{k}) + \overline{\mathbf{J}}^\dagger (\mathbf{k}) \, \overline{\mathbf{M}} (\mathbf{k}) \big) \\
        &= \frac{\mu_0}{c} \int _{\mathbb{R}^3} \mathrm{d}^3 k \, \frac{1}{ |\mathbf{k}|^2} \overline{\mathbf{M}}^* (\mathbf{k})  \cdot \overline{\mathbf{J}} (\mathbf{k}) + c.c. \\
        &= \frac{\mu _0}{c} \frac{1}{\sqrt{(2\pi)^3}}\int _{\mathbb{R}^3} \mathrm{d}^3 r \int _{\mathbb{R}^3} \mathrm{d}^3 r' \,\sqrt{\frac{\pi}{2}}\frac{1}{|\mathbf{r}-\mathbf{r}'|} {\mathbf{M}} ^* (\mathbf{r}) \cdot {\mathbf{J}}(\mathbf{r}') + c.c. \\
        &= \frac{\mu_0}{ 2 \pi c}\int _{\mathbb{R}^3} \mathrm{d}^3r \int _{\mathbb{R}^3} \mathrm{d}^3r' \, \frac{\mathbf{M} (\mathbf{r})\cdot  \mathbf{J} (\mathbf{r}') }{|\mathbf{r}-\mathbf{r}'|} \ .
    \end{aligned}
\end{equation}
\end{widetext}
In the transformation above, we have performed the following steps. 
From the form of $\Lambda_{JM}$ in Eq.~(\ref{eq:helicity_magnetic_k}) in the first line, we can see that the two terms in the parentheses are complex conjugates ($c.c.$), so we rewrite the expression accordingly, again visually separating the scalar function, $\overline{f}(\mathbf{k}) = \frac{1}{ |\mathbf{k}|^2}$, and the product of two vectors, $\overline{\mathbf{a}}^*(\mathbf{k}) =\overline{\mathbf{M}}^*(\mathbf{k})  $ and $\overline{\mathbf{b}}(\mathbf{k}) = \overline{\mathbf{J}} (\mathbf{k})$. 
Then we use Eqs.~(\ref{app_eq:generalized_parseval}) and~(\ref{app_eq:FT_1/k2}) to arrive at the real space formulation of the integral in the third line. 
The reality of $\mathbf{M}(\mathbf{r})$ and $\mathbf{J}(\mathbf{r})$ leads to the complex conjugate collapsing to the same expression, thus we get an additional factor of 2 from it. 
The final expression, coinciding with the one we use to evaluate the helicity in the main article, is given in the last row of the above equation.

\section{Equivalence of our Helicity Expressions to Magnetic Helicity}
\label{app:magnetic_helicity}
In the following, we will prove that Eq.~(\ref{eq:helicity_magnetic_k}) is equivalent to the magnetic helicity 
\begin{equation}
\label{app_eq:mag_hel_BA}
    \int _{\mathbb{R}^3} \mathrm{d}^3r \, \mathbf{B}(\mathbf{r})\cdot\mathbf{A}(\mathbf{r})
\end{equation}, only differing by a factor of $\mu_0 c$, as given in Eq.~(\ref{eq:helicity_equivalence}). 
With the vector potential given in Eq.~(\ref{eq:A(r)}) as
\begin{equation}
\begin{aligned}
    \label{eq:A(r)_appendix}
    \mathbf{A}(\mathbf{r}) = & \mathbf{A}_J(\mathbf{r}) + \mathbf{A}_M(\mathbf{r}) \\ 
    &=\frac{\mu_0}{4\pi} \int _{\mathbb{R}^3} \mathrm{d}^3r'\, \frac{\mathbf{J}(\mathbf{r'})}{|\mathbf{r}-\mathbf{r'}|} \\ &+  \frac{\mu_0}{4\pi} \int _{\mathbb{R}^3} \mathrm{d}^3r' \, \frac{\nabla' \times \mathbf{M}(\mathbf{r'})}{|\mathbf{r}-\mathbf{r'}|} \, ,
\end{aligned}
\end{equation}
where we will work separately with the two parts of the vector potential associated with the current and magnetization density, respectively, for easier handling. 
The magnetic field is then given by
\begin{equation} 
    \begin{aligned}
        \mathbf{B}(\mathbf{r}) &= \nabla \times \mathbf{A}(\mathbf{r}) \\
        &= \mathbf{B}_J(\mathbf{r}) + \mathbf{B}_M(\mathbf{r}) \, ,
    \end{aligned}
\end{equation}
which is also separated into two contributions. 
For the part associated with the current density, we just take the curl of the corresponding vector potential as
\begin{equation} \label{app_eq:B_J}
    \begin{aligned}
        \mathbf{B}_J(\mathbf{r}) &= \nabla \times \mathbf{A}_J(\mathbf{r}) \\
        &= \frac{\mu_0}{4\pi} \, \nabla \times\int _{\mathbb{R}^3} \mathrm{d}^3r'\, \frac{\mathbf{J}(\mathbf{r'})}{|\mathbf{r}-\mathbf{r'}|} \\
        &= \frac{\mu_0}{4\pi} \int _{\mathbb{R}^3} \mathrm{d}^3r' \, \frac{\mathbf{J}(\mathbf{r}') \times (\mathbf{r}-\mathbf{r}')}{|\mathbf{r}-\mathbf{r}'|^3} \, ,\\ \vspace{0.1mm}
    \end{aligned}
\end{equation}
while that associated with the magnetization density is simply
\begin{equation}
\label{app_eq:B=M}
    \mathbf{B}_M(\mathbf{r}) = \mu _0 \mathbf{M}(\mathbf{r}) \, ,
\end{equation}
as follows from Eq.~(\ref{eq:B_k}) for a transverse $\mathbf{M}(\mathbf{r})$. 

We have already mentioned the common factor of $\mu_0 c$ between all terms in $\braket{\Lambda}$ and the magnetic helicity, Eq.~(\ref{app_eq:mag_hel_BA}) in the main article (see Eq. (\ref{eq:helicity_equivalence})). 
We will hence not consider it further here. 

Now, let us go through the different terms one by one. 
The pure magnetization term (the last line in Eq.~(\ref{app_eq:lambdaM_McurlM_k_to_r})) is straightforward to get by substituting the last line of Eq.~(\ref{eq:A(r)_appendix}) and Eq.~(\ref{app_eq:B=M}) into Eq.~(\ref{app_eq:mag_hel_BA}). 
The pure current term (the last line in Eq.~(\ref{app_eq:lambdaJ_k_to_r})) needs equivalent steps: we substitute the second line in Eq.~(\ref{eq:A(r)_appendix}) and the last line in Eq.~(\ref{app_eq:B_J}) into Eq.~(\ref{app_eq:mag_hel_BA}). 
Finally, the mixed term (the last line in Eq.~(\ref{app_eq:lambdaJM_k_to_r})), when performing a similar substitution as for the other two terms, acquires two separate terms: $\int _{\mathbb{R}^3} \mathrm{d}^3r \, \mathbf{B}_M(\mathbf{r})\cdot\mathbf{A}_J(\mathbf{r})$ and $\int _{\mathbb{R}^3} \mathrm{d}^3r \, \mathbf{B}_J(\mathbf{r})\cdot\mathbf{A}_M(\mathbf{r})$. 
The first of those comes out to be half of our $\Lambda_{JM}$ directly when substituting the second line in Eq.~(\ref{eq:A(r)_appendix}) and Eq.~(\ref{app_eq:B=M}). 
The second term can be transformed to also yield one half of $\Lambda_{JM}$. 
This involves a series of vector calculus identities to handle the derivatives and integrals. 
We opt to not show this lengthier derivation here, as we know already that the two terms in $\Lambda_{JM}$ are complex conjugates of each other and since all quantities are real, this simply leads to a factor of 2 in the term coming out of $\int _{\mathbb{R}^3} \mathrm{d}^3r \, \mathbf{B}_M(\mathbf{r})\cdot\mathbf{A}_J(\mathbf{r})$. 
With this, we have shown that our helicity expressions are equivalent to the magnetic helicity coming from magnetization and current densities.

\bibliography{references}

@article{knapmanNumericalCalculationHopf2025,
	title = {Numerical calculation of the {Hopf} index for three-dimensional magnetic textures},
	volume = {111},
	issn = {2469-9950, 2469-9969},
	url = {https://link.aps.org/doi/10.1103/PhysRevB.111.134408},
	doi = {10.1103/PhysRevB.111.134408},
	number = {13},
	urldate = {2026-11-25},
	journal = {Physical Review B},
	author = {Knapman, R. and Azhar, M. and Pignedoli, A. and Gallard, L. and Hertel, R. and Leliaert, J. and Everschor-Sitte, K.},
	month = apr,
	year = {2025},
	pages = {134408},
}

@article{kentCreationObservationHopfions2021,
	title = {Creation and observation of {Hopfions} in magnetic multilayer systems},
	volume = {12},
	issn = {2041-1723},
	url = {https://www.nature.com/articles/s41467-021-21846-5},
	doi = {10.1038/s41467-021-21846-5},
	number = {1},
	urldate = {2026-11-25},
	journal = {Nature Communications},
	author = {Kent, Noah and Reynolds, Neal and Raftrey, David and Campbell, Ian T. G. and Virasawmy, Selven and Dhuey, Scott and Chopdekar, Rajesh V. and Hierro-Rodriguez, Aurelio and Sorrentino, Andrea and Pereiro, Eva and Ferrer, Salvador and Hellman, Frances and Sutcliffe, Paul and Fischer, Peter},
	month = mar,
	year = {2021},
	pages = {1562},
}

@phdthesis{poulikakosChiralLightMatterInteractions2018,
	title = {Chiral {Light}-{Matter} {Interactions} in the {Near} and {Far} {Field}},
	copyright = {http://rightsstatements.org/page/InC-NC/1.0/, info:eu-repo/semantics/openAccess},
	url = {http://hdl.handle.net/20.500.11850/296342},
	doi = {10.3929/ETHZ-B-000296342},
	urldate = {2026-11-25},
	school = {ETH Zurich},
	author = {Poulikakos, Lisa V.},
	collaborator = {{Norris, David J.} and {Genet, Cyriaque} and {Kuipers, Laurens K.} and {Novotny, Lukas}},
	year = {2018},
}

@inproceedings{poulikakos_optChirality_inhomSpace,
	booktitle = {Photonic and phononic properties of engineered nanostructures {VI}},
	title = {Time-harmonic optical chirality in inhomogeneous space},
	url = {http://proceedings.spiedigitallibrary.org/proceeding.aspx?doi=10.1117/12.2209551},
	doi = {10.1117/12.2209551},
	urldate = {2026-11-25},
	author = {Gutsche, Philipp and Poulikakos, Lisa V. and Hammerschmidt, Martin and Burger, Sven and Schmidt, Frank},
	month = mar,
	year = {2016},
	pages = {97560X},
}

@article{poulikakos_OpticalChiralityFlux2016,
	title = {Optical chirality flux as a useful far-field probe of chiral near fields},
	volume = {3},
	issn = {2330-4022},
	url = {https://doi.org/10.1021/acsphotonics.6b00201},
	doi = {10.1021/acsphotonics.6b00201},
	number = {9},
	journal = {ACS Photonics},
	author = {Poulikakos, Lisa V. and Gutsche, Philipp and McPeak, Kevin M. and Burger, Sven and Niegemann, Jens and Hafner, Christian and Norris, David J.},
	month = sep,
	year = {2016},
	pages = {1619--1625},
}

@article{livesayElectromagneticFieldsInduced1974,
	title = {Electromagnetic {Fields} {Induced} {Inside} {Arbitrarily} {Shaped} {Biological} {Bodies}},
	volume = {22},
	copyright = {https://ieeexplore.ieee.org/Xplorehelp/downloads/license-information/IEEE.html},
	issn = {0018-9480},
	url = {http://ieeexplore.ieee.org/document/1128475/},
	doi = {10.1109/TMTT.1974.1128475},
	language = {en},
	number = {12},
	urldate = {2026-11-02},
	journal = {IEEE Transactions on Microwave Theory and Techniques},
	author = {Livesay, D.E. and Chen, K.},
	month = dec,
	year = {1974},
	pages = {1273--1280},
}

@article{franzkeTURBOMOLETodayTomorrow2023,
	title = {{TURBOMOLE}: {Today} and {Tomorrow}},
	volume = {19},
	copyright = {https://creativecommons.org/licenses/by/4.0/},
	issn = {1549-9618, 1549-9626},
	shorttitle = {{TURBOMOLE}},
	url = {https://pubs.acs.org/doi/10.1021/acs.jctc.3c00347},
	doi = {10.1021/acs.jctc.3c00347},
	language = {en},
	number = {20},
	urldate = {2026-10-15},
	journal = {Journal of Chemical Theory and Computation},
	author = {Franzke, Yannick J. and Holzer, Christof and Andersen, Josefine H. and Begušić, Tomislav and Bruder, Florian and Coriani, Sonia and Della Sala, Fabio and Fabiano, Eduardo and Fedotov, Daniil A. and Fürst, Susanne and Gillhuber, Sebastian and Grotjahn, Robin and Kaupp, Martin and Kehry, Max and Krstić, Marjan and Mack, Fabian and Majumdar, Sourav and Nguyen, Brian D. and Parker, Shane M. and Pauly, Fabian and Pausch, Ansgar and Perlt, Eva and Phun, Gabriel S. and Rajabi, Ahmadreza and Rappoport, Dmitrij and Samal, Bibek and Schrader, Tim and Sharma, Manas and Tapavicza, Enrico and Treß, Robert S. and Voora, Vamsee and Wodyński, Artur and Yu, Jason M. and Zerulla, Benedikt and Furche, Filipp and Hättig, Christof and Sierka, Marek and Tew, David P. and Weigend, Florian},
	month = oct,
	year = {2023},
	pages = {6859--6890},
}

@article{sundholmCurrentDensityMolecular2021,
	title = {Current density and molecular magnetic properties},
	volume = {57},
	issn = {1359-7345},
	url = {https://doi.org/10.1039/d1cc03350f},
	doi = {10.1039/d1cc03350f},
	number = {93},
	urldate = {2026-09-09},
	journal = {Chemical Communications},
	author = {Sundholm, Dage and Dimitrova, Maria and Berger, Raphael J. F.},
	month = dec,
	year = {2021},
	pages = {12362--12378},
}

@article{tellgrenDensityfunctionalTheoryInternal2018,
	title = {Density-functional theory for internal magnetic fields},
	volume = {97},
	url = {https://link.aps.org/doi/10.1103/PhysRevA.97.012504},
	doi = {10.1103/PhysRevA.97.012504},
	number = {1},
	urldate = {2026-09-09},
	journal = {Physical Review A},
	publisher = {American Physical Society},
	author = {Tellgren, Erik I.},
	month = jan,
	year = {2018},
	pages = {012504},
}

@article{holzerCurrentDensityFunctional2022,
	title = {Current density functional framework for spin–orbit coupling},
	volume = {157},
	issn = {0021-9606},
	url = {https://doi.org/10.1063/5.0122394},
	doi = {10.1063/5.0122394},
	number = {20},
	urldate = {2026-09-09},
	journal = {The Journal of Chemical Physics},
	author = {Holzer, Christof and Franzke, Yannick J. and Pausch, Ansgar},
	month = nov,
	year = {2022},
	pages = {204102},
}

@article{bergerRigorousNewLimits1984,
	title = {Rigorous new limits on magnetic helicity dissipation in the solar corona},
	volume = {30},
	issn = {0309-1929},
	url = {https://doi.org/10.1080/03091928408210078},
	doi = {10.1080/03091928408210078},
	number = {1-2},
	urldate = {2026-09-08},
	journal = {Geophysical \& Astrophysical Fluid Dynamics},
	publisher = {Taylor \& Francis},
	author = {Berger, Mitchell A.},
	month = sep,
	year = {1984},
	pages = {79--104},
}

@article{finnMagneticHelicityWhat1985,
	title = {Magnetic helicity: {What} is it and what is it good for?},
	volume = {9},
	shorttitle = {Magnetic helicity},
	url = {https://ui.adsabs.harvard.edu/abs/1985CoPPC...9..111F},
	urldate = {2026-09-09},
	journal = {Comments on Plasma Physics and Controlled Fusion},
	author = {Finn, John M. and Antonsen, Jr., Thomas M.},
	month = may,
	year = {1985},
	pages = {111--126},
}

@book{arfkenMathematicalMethodsPhysicists2013,
	address = {Amsterdam Boston},
	edition = {7th ed},
	title = {Mathematical methods for physicists: a comprehensive guide},
	isbn = {978-0-12-384654-9},
	shorttitle = {Mathematical methods for physicists},
	publisher = {Elsevier},
	author = {Arfken, George and Weber, Hans-Jurgen and Harris, Frank E.},
	year = {2013},
}

@book{bellanFundamentalsPlasmaPhysics2006,
	edition = {1},
	title = {Fundamentals of {Plasma} {Physics}},
	copyright = {https://www.cambridge.org/core/terms},
	isbn = {978-0-521-82116-2 978-0-521-52800-9 978-0-511-80718-3},
	url = {https://www.cambridge.org/core/product/identifier/9780511807183/type/book},
	doi = {10.1017/CBO9780511807183},
	urldate = {2026-09-08},
	publisher = {Cambridge University Press},
	author = {Bellan, Paul M.},
	month = apr,
	year = {2006},
}

@book{brownMagneticHelicitySpace2010,
	address = {Washington, DC},
	series = {Geophysical monograph},
	title = {Magnetic helicity in space and laboratory plasmas},
	isbn = {978-0-87590-094-0 978-1-118-66447-6},
	number = {111},
	publisher = {American Geophysical Union},
	editor = {Brown, Michael Riley and Canfield, Richard C. and Pevtsov, Alexei A.},
	year = {2010},
    doi={10.1029/GM111},
}

@INPROCEEDINGS{kusanoStudyMagneticHelicity2004,
       author = {{Kusano}, K. and {Maeshiro}, T. and {Yokoyama}, T. and {Sakurai}, T.},
        title = "{Study of Magnetic Helicity in the Solar Corona}",
    booktitle = {The Solar-B Mission and the Forefront of Solar Physics},
         year = 2004,
       volume = {325},
        month = dec,
        pages = {175},
       adsurl = {https://scixplorer.org/abs/2004ASPC..325..175K},
      isbn    = {978-1-58381-260-0}
}

@article{jiConservationMagneticHelicity1995,
	title = {Conservation of {Magnetic} {Helicity} during {Plasma} {Relaxation}},
	volume = {74},
	url = {https://link.aps.org/doi/10.1103/PhysRevLett.74.2945},
	doi = {10.1103/PhysRevLett.74.2945},
	number = {15},
	urldate = {2026-09-08},
	journal = {Physical Review Letters},
	publisher = {American Physical Society},
	author = {Ji, H. and Prager, S. C. and Sarff, J. S.},
	month = apr,
	year = {1995},
	pages = {2945--2948},
}

@article{yeatesEvolutionFieldLine2021,
	title = {Evolution of field line helicity in magnetic relaxation},
	volume = {28},
	issn = {1070-664X},
	url = {https://doi.org/10.1063/5.0059756},
	doi = {10.1063/5.0059756},
	number = {8},
	urldate = {2026-09-08},
	journal = {Physics of Plasmas},
	author = {Yeates, A. R. and Russell, A. J. B. and Hornig, G.},
	month = aug,
	year = {2021},
	pages = {082904},
}

@article{boozerMagneticHelicityDynamos1993,
	title = {Magnetic helicity and dynamos},
	volume = {5},
	issn = {0899-8221},
	url = {https://doi.org/10.1063/1.860761},
	doi = {10.1063/1.860761},
	number = {7},
	urldate = {2026-09-08},
	journal = {Physics of Fluids B: Plasma Physics},
	author = {Boozer, Allen H.},
	month = jul,
	year = {1993},
	pages = {2271--2277},
}

@incollection{ImplicationsMagneticHelicity,
	address = {Washington, D. C.},
	series = {Geophysical {Monograph} {Series}},
	title = {Implications of {Magnetic} {Helicity} {Conservation}},
	isbn = {978-1-118-66447-6 978-0-87590-094-0},
	url = {http://doi.wiley.com/10.1029/GM111p0011},
	doi = {10.1029/GM111p0011},
	urldate = {2026-11-25},
	booktitle = {Magnetic {Helicity} in {Space} and {Laboratory} {Plasmas}},
	publisher = {American Geophysical Union},
	author = {Boozer, Allen H.},
	editor = {Brown, Michael R. and Canfield, Richard C. and Pevtsov, Alexei A.},
	month = mar,
	year = {2013},
	pages = {11--16},
}

@article{sutcliffeHopfionsChiralMagnets2018,
	title = {Hopfions in chiral magnets},
	volume = {51},
	issn = {1751-8121},
	url = {https://doi.org/10.1088/1751-8121/aad521},
	doi = {10.1088/1751-8121/aad521},
	number = {37},
	urldate = {2026-09-07},
	journal = {Journal of Physics A: Mathematical and Theoretical},
	publisher = {IOP Publishing},
	author = {Sutcliffe, Paul},
	month = aug,
	year = {2018},
	pages = {375401},
}

@article{taylorRelaxationToroidalPlasma1974,
	title = {Relaxation of {Toroidal} {Plasma} and {Generation} of {Reverse} {Magnetic} {Fields}},
	volume = {33},
	url = {https://link.aps.org/doi/10.1103/PhysRevLett.33.1139},
	doi = {10.1103/PhysRevLett.33.1139},
	number = {19},
	urldate = {2026-09-08},
	journal = {Physical Review Letters},
	publisher = {American Physical Society},
	author = {Taylor, J. B.},
	month = nov,
	year = {1974},
	pages = {1139--1141},
}

@article{kosevichMagneticSolitons1990,
	title = {Magnetic {Solitons}},
	volume = {194},
	issn = {0370-1573},
	url = {https://www.sciencedirect.com/science/article/pii/037015739090130T},
	doi = {10.1016/0370-1573(90)90130-T},
	number = {3},
	urldate = {2026-09-07},
	journal = {Physics Reports},
	author = {Kosevich, A. M. and Ivanov, B. A. and Kovalev, A. S.},
	month = oct,
	year = {1990},
	pages = {117--238},
}

@article{shajuHopfionsScrewChiral2026,
	title = {Hopfions in screw chiral magnets},
	volume = {14},
	issn = {2166-532X},
	url = {https://doi.org/10.1063/5.0327533},
	doi = {10.1063/5.0327533},
	number = {5},
	urldate = {2026-09-07},
	journal = {APL Materials},
	author = {Shaju, Sandra Chulliparambil and Azhar, Maria and Everschor-Sitte, Karin},
	month = may,
	year = {2026},
	pages = {051111},
}

@article{sutcliffeSkyrmionKnotsFrustrated2017,
	title = {Skyrmion {Knots} in {Frustrated} {Magnets}},
	volume = {118},
	doi = {10.1103/PhysRevLett.118.247203},
	number = {24},
	journal = {Physical Review Letters},
	author = {Sutcliffe, Paul},
	year = {2017},
}

@article{brandenburgCriticalRoleMagnetic2009,
	title = {The critical role of magnetic helicity in astrophysical large-scale dynamos},
	volume = {51},
	issn = {0741-3335},
	url = {https://doi.org/10.1088/0741-3335/51/12/124043},
	doi = {10.1088/0741-3335/51/12/124043},
	number = {12},
	urldate = {2026-09-03},
	journal = {Plasma Physics and Controlled Fusion},
	author = {Brandenburg, Axel},
	month = nov,
	year = {2009},
	pages = {124043},
}

@article{blackmanMagneticHelicityLarge2015,
	title = {Magnetic {Helicity} and {Large} {Scale} {Magnetic} {Fields}: {A} {Primer}},
	volume = {188},
	issn = {1572-9672},
	shorttitle = {Magnetic {Helicity} and {Large} {Scale} {Magnetic} {Fields}},
	url = {https://doi.org/10.1007/s11214-014-0038-6},
	doi = {10.1007/s11214-014-0038-6},
	number = {1},
	urldate = {2026-09-03},
	journal = {Space Science Reviews},
	author = {Blackman, Eric G.},
	month = may,
	year = {2015},
	pages = {59--91},
}

@article{moffattHelicity2017,
	title = {Helicity},
	volume = {346},
	issn = {1873-7234},
	url = {https://comptes-rendus.academie-sciences.fr/mecanique/articles/10.1016/j.crme.2017.12.002/},
	doi = {10.1016/j.crme.2017.12.002},
	number = {3},
	urldate = {2026-09-03},
	journal = {Comptes Rendus. Mécanique},
	author = {Moffatt, H. Keith},
	month = dec,
	year = {2017},
	pages = {165--169},
}

@article{elsasserHydromagneticDynamoTheory1956,
	title = {Hydromagnetic {Dynamo} {Theory}},
	volume = {28},
	url = {https://link.aps.org/doi/10.1103/RevModPhys.28.135},
	doi = {10.1103/RevModPhys.28.135},
	number = {2},
	urldate = {2026-09-03},
	journal = {Reviews of Modern Physics},
	publisher = {American Physical Society},
	author = {Elsasser, Walter M.},
	month = apr,
	year = {1956},
	pages = {135--163},
}

@article{soltanitehraniGalileanInvarianceMean2023,
	title = {On {Galilean} invariance of mean kinetic helicity},
	volume = {35},
	issn = {1070-6631},
	url = {https://doi.org/10.1063/5.0178926},
	doi = {10.1063/5.0178926},
	number = {12},
	urldate = {2026-09-03},
	journal = {Physics of Fluids},
	author = {Soltani Tehrani, Dina and Aluie, H.},
	month = dec,
	year = {2023},
	pages = {121701},
}

@book{kuzanyanHelicitiesGeophysicsAstrophysics2024,
	address = {Washington, DC},
	series = {Geophysical monograph series},
	title = {Helicities in geophysics, astrophysics, and beyond},
	isbn = {978-1-119-84168-5 978-1-119-84171-5 978-1-119-84169-2 978-1-119-84170-8},
	doi = {10.1002/9781119841715},
	number = {283},
	publisher = {American Geophysical Union},
	editor = {Kuzanyan, Kirill and Yokoi, Nobumitsu and Georgoulis, Manolis and Stepanov, Rodion},
	year = {2024},
}

@article{poulikakosOpticalHelicityOptical2019,
	title = {Optical {Helicity} and {Optical} {Chirality} in {Free} {Space} and in the {Presence} of {Matter}},
	volume = {11},
	copyright = {http://creativecommons.org/licenses/by/3.0/},
	issn = {2073-8994},
	url = {https://www.mdpi.com/2073-8994/11/9/1113},
	doi = {10.3390/sym11091113},
	number = {9},
	urldate = {2026-09-03},
	journal = {Symmetry},
	publisher = {Multidisciplinary Digital Publishing Institute},
	author = {Poulikakos, Lisa V. and Dionne, Jennifer A. and García-Etxarri, Aitzol},
	month = sep,
	year = {2019},
	pages = {1113},
}

@article{cameronChiralityAngularMomentum2017,
	title = {Chirality and the angular momentum of light},
	volume = {375},
	issn = {1364-503X},
	url = {https://doi.org/10.1098/rsta.2015.0433},
	doi = {10.1098/rsta.2015.0433},
	number = {2087},
	urldate = {2026-09-03},
	journal = {Philosophical Transactions of the Royal Society A: Mathematical, Physical and Engineering Sciences},
	author = {Cameron, Robert P. and Götte, Jörg B. and Barnett, Stephen M. and Yao, Alison M.},
	month = feb,
	year = {2017},
	pages = {20150433},
}

@article{lipkinExistenceNewConservation1964,
	title = {Existence of a {New} {Conservation} {Law} in {Electromagnetic} {Theory}},
	volume = {5},
	issn = {0022-2488},
	url = {https://doi.org/10.1063/1.1704165},
	doi = {10.1063/1.1704165},
	number = {5},
	urldate = {2026-09-03},
	journal = {Journal of Mathematical Physics},
	author = {Lipkin, Daniel M.},
	month = may,
	year = {1964},
	pages = {696--700},
}

@article{tangOpticalChiralityIts2010,
	title = {Optical {Chirality} and {Its} {Interaction} with {Matter}},
	volume = {104},
	url = {https://link.aps.org/doi/10.1103/PhysRevLett.104.163901},
	doi = {10.1103/PhysRevLett.104.163901},
	number = {16},
	urldate = {2026-09-03},
	journal = {Physical Review Letters},
	publisher = {American Physical Society},
	author = {Tang, Yiqiao and Cohen, Adam E.},
	month = apr,
	year = {2010},
	pages = {163901},
}

@article{liningerChiralityLightMatter2023,
	title = {Chirality in {Light}–{Matter} {Interaction}},
	volume = {35},
	copyright = {© 2022 The Authors. Advanced Materials published by Wiley-VCH GmbH},
	issn = {1521-4095},
	url = {https://onlinelibrary.wiley.com/doi/abs/10.1002/adma.202107325},
	doi = {10.1002/adma.202107325},
	number = {34},
	urldate = {2026-09-03},
	journal = {Advanced Materials},
	author = {Lininger, Andrew and Palermo, Giovanna and Guglielmelli, Alexa and Nicoletta, Giuseppe and Goel, Madhav and Hinczewski, Michael and Strangi, Giuseppe},
	year = {2023},
	pages = {2107325},
}

@article{fernandez-corbatonHelicityAngularMomentum2012,
	title = {Helicity and angular momentum: {A} symmetry-based framework for the study of light-matter interactions},
	volume = {86},
	shorttitle = {Helicity and angular momentum},
	url = {https://link.aps.org/doi/10.1103/PhysRevA.86.042103},
	doi = {10.1103/PhysRevA.86.042103},
	number = {4},
	urldate = {2026-09-03},
	journal = {Physical Review A},
	publisher = {American Physical Society},
	author = {Fernandez-Corbaton, Ivan and Zambrana-Puyalto, Xavier and Molina-Terriza, Gabriel},
	month = oct,
	year = {2012},
	pages = {042103},
}

@article{fernandez-corbatonConformallyInvariantDerivation2019,
	title = {A {Conformally} {Invariant} {Derivation} of {Average} {Electromagnetic} {Helicity}},
	volume = {11},
	copyright = {http://creativecommons.org/licenses/by/3.0/},
	issn = {2073-8994},
	url = {https://www.mdpi.com/2073-8994/11/11/1427},
	doi = {10.3390/sym11111427},
	number = {11},
	urldate = {2026-09-02},
	journal = {Symmetry},
	publisher = {Multidisciplinary Digital Publishing Institute},
	author = {Fernandez-Corbaton, Ivan},
	month = nov,
	year = {2019},
	pages = {1427},
}

@article{afanasievHelicityFreeElectromagnetic1996,
	title = {The helicity of the free electromagnetic field and its physical meaning},
	volume = {109},
	issn = {1826-9869},
	url = {https://doi.org/10.1007/BF02731014},
	doi = {10.1007/BF02731014},
	number = {3},
	urldate = {2026-09-02},
	journal = {Il Nuovo Cimento A (1965-1970)},
	author = {Afanasiev, G. N. and Stepanovsky, Yu. P.},
	month = mar,
	year = {1996},
	pages = {271--279},
}

@article{truebaElectromagneticHelicity1996,
	title = {The electromagnetic helicity},
	volume = {17},
	issn = {0143-0807},
	url = {https://doi.org/10.1088/0143-0807/17/3/008},
	doi = {10.1088/0143-0807/17/3/008},
	number = {3},
	urldate = {2026-09-02},
	journal = {European Journal of Physics},
	author = {Trueba, José L. and Rañada, Antonio F.},
	month = may,
	year = {1996},
	pages = {141},
}

@article{drummondDualSymmetricLagrangians1999,
	title = {Dual symmetric {Lagrangians} and conservation laws},
	volume = {60},
	url = {https://link.aps.org/doi/10.1103/PhysRevA.60.R3331},
	doi = {10.1103/PhysRevA.60.R3331},
	number = {5},
	urldate = {2026-09-02},
	journal = {Physical Review A},
	publisher = {American Physical Society},
	author = {Drummond, P. D.},
	month = nov,
	year = {1999},
	pages = {R3331--R3334},
}

@article{bialynicki-birulaNoteHelicity1981,
	title = {A note on helicity},
	volume = {22},
	issn = {0022-2488},
	url = {https://doi.org/10.1063/1.524828},
	doi = {10.1063/1.524828},
	number = {11},
	urldate = {2026-09-02},
	journal = {Journal of Mathematical Physics},
	author = {Bialynicki‐Birula, Iwo and Newman, E. T. and Porter, J. and Winicour, J. and Lukacs, B. and Perjes, Z. and Sebestyen, A.},
	month = nov,
	year = {1981},
	pages = {2530--2532},
}

@article{deserDualityTransformationsAbelian1976,
	title = {Duality transformations of {Abelian} and non-{Abelian} gauge fields},
	volume = {13},
	doi = {10.1103/PhysRevD.13.1592},
	number = {6},
	journal = {Physical Review D},
	author = {Deser, Stanley},
	year = {1976},
	pages = {1592--1597},
}

@article{calkinInvariancePropertyFree1965,
	title = {An {Invariance} {Property} of the {Free} {Electromagnetic} {Field}},
	volume = {33},
	issn = {0002-9505},
	url = {https://doi.org/10.1119/1.1971089},
	doi = {10.1119/1.1971089},
	number = {11},
	urldate = {2026-09-02},
	journal = {American Journal of Physics},
	author = {Calkin, M. G.},
	month = nov,
	year = {1965},
	pages = {958--960},
}

@book{strattonElectromagneticTheory2007,
	address = {Piscataway, NJ},
	series = {{IEEE} {Press} series on electromagnetic wave theory},
	title = {Electromagnetic theory},
	isbn = {978-0-470-13153-4 978-1-119-13464-0 978-1-60119-510-4},
	doi = {10.1002/9781119134640},
	publisher = {IEEE Press},
	author = {Stratton, Julius Adams},
	collaborator = {{IEEE Antennas and Propagation Society}},
	year = {2007},
}

@article{azhar3DMagneticTextures2026,
	title = {{3D} magnetic textures with mixed topology: {Unlocking} the tunable {Hopf} index},
	volume = {21},
	issn = {2542-4653},
	shorttitle = {{3D} magnetic textures with mixed topology},
	url = {https://scipost.org/10.21468/SciPostPhys.21.1.023},
	doi = {10.21468/SciPostPhys.21.1.023},
	number = {1},
	urldate = {2026-08-31},
	journal = {SciPost Physics},
	author = {Azhar, Maria and Shaju, Sandra C. and Knapman, Ross and Pignedoli, Alessandro and Everschor-Sitte, Karin},
	month = jul,
	year = {2026},
	pages = {023},
}

@article{bergerTopologicalPropertiesMagnetic1984,
	title = {The topological properties of magnetic helicity},
	volume = {147},
	issn = {1469-7645, 0022-1120},
	url = {https://www.cambridge.org/core/journals/journal-of-fluid-mechanics/article/abs/topological-properties-of-magnetic-helicity/CBF3020A84DF9DFA4894162430595777},
	doi = {10.1017/S0022112084002019},
	urldate = {2026-09-01},
	journal = {Journal of Fluid Mechanics},
	author = {Berger, Mitchell A. and Field, George B.},
	month = oct,
	year = {1984},
	pages = {133--148},
}

@article{hopfUberAbbildungenDreidimensionalen1931,
	title = {Über die {Abbildungen} der dreidimensionalen {Sphäre} auf die {Kugelfläche}},
	volume = {104},
	issn = {0025-5831; 1432-1807/e},
	url = {https://eudml.org/doc/159489},
	urldate = {2026-08-24},
	journal = {Mathematische Annalen},
	author = {Hopf, H.},
	year = {1931},
	pages = {637--665},
}

@article{moffattDegreeKnottednessTangled1969,
	title = {The degree of knottedness of tangled vortex lines},
	volume = {35},
	issn = {1469-7645, 0022-1120},
	url = {https://www.cambridge.org/core/journals/journal-of-fluid-mechanics/article/abs/degree-of-knottedness-of-tangled-vortex-lines/798DFB897A75D97CE3081F4C1DA970F0},
	doi = {10.1017/S0022112069000991},
	number = {1},
	urldate = {2026-08-24},
	journal = {Journal of Fluid Mechanics},
	author = {Moffatt, H. K.},
	month = jan,
	year = {1969},
	pages = {117--129},
}

@article{moffattHelicityLaminarTurbulent1992,
	title = {Helicity in {Laminar} and {Turbulent} {Flow}},
	volume = {24},
	issn = {1545-4479},
	url = {https://www.annualreviews.org/content/journals/10.1146/annurev.fl.24.010192.001433},
	doi = {10.1146/annurev.fl.24.010192.001433},
	urldate = {2026-09-01},
	journal = {Annual Review of Fluid Mechanics},
	publisher = {Annual Reviews},
	author = {Moffatt, H. K. and Tsinober, A.},
	year = {1992},
	pages = {281--312},
}

@article{moreauConstantesDunIlot1961,
	title = {Constantes d'un îlot tourbillonnaire en fluide parfait barotrope},
	volume = {252},
	url = {https://hal.science/hal-01865239},
	urldate = {2026-09-01},
	journal = {Comptes rendus hebdomadaires des séances de l'Académie des sciences},
	publisher = {Gauthier-Villars},
	author = {Moreau, Jean Jacques},
	year = {1961},
	pages = {2810--2812},
}

@article{whiteheadExpressionHopfsInvariant1947,
	title = {An {Expression} of {Hopf}'s {Invariant} as an {Integral}*},
	volume = {33},
	url = {https://www.pnas.org/doi/abs/10.1073/pnas.33.5.117},
	doi = {10.1073/pnas.33.5.117},
	number = {5},
	urldate = {2026-09-01},
	journal = {Proceedings of the National Academy of Sciences},
	publisher = {Proceedings of the National Academy of Sciences},
	author = {Whitehead, J. H. C.},
	month = may,
	year = {1947},
	pages = {117--123},
}

@article{guslienkoEmergentMagneticField2023,
	title = {Emergent magnetic field and vector potential of the toroidal magnetic hopfions},
	volume = {174},
	issn = {0960-0779},
	url = {https://www.sciencedirect.com/science/article/pii/S0960077923007415},
	doi = {10.1016/j.chaos.2023.113840},
	urldate = {2026-08-31},
	journal = {Chaos, Solitons \& Fractals},
	author = {Guslienko, Konstantin Y.},
	month = sep,
	year = {2023},
	pages = {113840},
}

@article{kuznetsovTopologicalMeaningCanonical1980,
	title = {On the topological meaning of canonical {Clebsch} variables},
	volume = {77},
	issn = {0375-9601},
	url = {https://www.sciencedirect.com/science/article/pii/0375960180906271},
	doi = {10.1016/0375-9601(80)90627-1},
	number = {1},
	urldate = {2026-08-24},
	journal = {Physics Letters A},
	author = {Kuznetsov, E. A. and Mikhailov, A. V.},
	month = apr,
	year = {1980},
	pages = {37--38},
}

@incollection{bialynicki-birulaPhotonWaveFunction1996,
	address = {Amsterdam},
	title = {Photon {Wave} {Function}},
	volume = {36},
	copyright = {https://www.elsevier.com/tdm/userlicense/1.0/},
	isbn = {978-0-444-82530-8},
	url = {https://linkinghub.elsevier.com/retrieve/pii/S0079663808703160},
	doi = {10.1016/S0079-6638(08)70316-0},
	urldate = {2026-06-10},
	booktitle = {Progress in {Optics}},
	publisher = {Elsevier},
	author = {Bialynicki-Birula, Iwo},
	year = {1996},
	pages = {245--294},
}

@article{grossNormInvarianceMassZero1964,
	title = {Norm {Invariance} of {Mass}-{Zero} {Equations} under the {Conformal} {Group}},
	volume = {5},
	issn = {0022-2488, 1089-7658},
	url = {https://pubs.aip.org/jmp/article/5/5/687/394316/Norm-Invariance-of-Mass-Zero-Equations-under-the},
	doi = {10.1063/1.1704164},
	number = {5},
	urldate = {2026-06-10},
	journal = {Journal of Mathematical Physics},
	author = {Gross, Leonard},
	month = may,
	year = {1964},
	pages = {687--695},
}

@article{fernandez-corbatonTotalHelicityElectromagnetic2021,
	title = {Total helicity of electromagnetic fields and matter},
	volume = {103},
	issn = {2469-9950, 2469-9969},
	url = {https://link.aps.org/doi/10.1103/PhysRevB.103.054406},
	doi = {10.1103/PhysRevB.103.054406},
	number = {5},
	urldate = {2026-06-10},
	journal = {Physical Review B},
	author = {Fernandez-Corbaton, Ivan},
	month = feb,
	year = {2021},
	pages = {054406},
}

@article{fernandez-corbatonScalarProductComputing2023,
	title = {A {Scalar} {Product} for {Computing} {Fundamental} {Quantities} in {Matter}},
	volume = {15},
	issn = {2073-8994},
	url = {https://www.mdpi.com/2073-8994/15/10/1839},
	doi = {10.3390/sym15101839},
	number = {10},
	urldate = {2026-06-10},
	journal = {Symmetry},
	author = {Fernandez-Corbaton, Ivan and Vavilin, Maxim},
	month = sep,
	year = {2023},
	pages = {1839},
}

@article{woltjerTheoremForcefreeMagnetic1958,
	title = {A theorem on force-free magnetic fields},
	volume = {44},
	url = {https://www.pnas.org/doi/abs/10.1073/pnas.44.6.489},
	doi = {10.1073/pnas.44.6.489},
	number = {6},
	urldate = {2026-06-10},
	journal = {Proceedings of the National Academy of Sciences},
	publisher = {Proceedings of the National Academy of Sciences},
	author = {Woltjer, L.},
	month = jun,
	year = {1958},
	pages = {489--491},
}

@Article{	  Pausch.Holzer:Linear.2022,
  author	= {Pausch, Ansgar and Holzer, Christof},
  title		= {Linear Response of Current-Dependent Density Functional
		  Approximations in Magnetic Fields},
  journal	= {J. Phys. Chem. Lett.},
  volume	= {13},
  number	= {19},
  pages		= {4335--4341},
  year		= {2022},
  doi		= {10.1021/acs.jpclett.2c01082},
  url		= { https://doi.org/10.1021/acs.jpclett.2c01082}
}

@Article{	  Tellgren.Teale.ea:Non-perturbative.2014,
  author	= {Tellgren, E. I. and Teale, A. M. and Furness, J. W. and
		  Lange, K. K. and Ekstr\"om, U. and Helgaker, T.},
  title		= {Non-perturbative calculation of molecular magnetic
		  properties within current-density functional theory},
  journal	= {J. Chem. Phys.},
  volume	= {140},
  pages		= {034101},
  year		= {2014},
  doi		= {10.1063/1.4861427},
  url		= { https://doi.org/10.1063/1.4861427}
}

@Article{	  Van-Wullen:Spin.2002,
  author	= {Van Wüllen, Christoph},
  title		= {Spin densities in two-component relativistic density
		  functional calculations: Noncollinear versus collinear
		  approach},
  journal	= {J. Comput. Chem.},
  volume	= {23},
  number	= {8},
  pages		= {779-785},
  doi		= {https://doi.org/10.1002/jcc.10043},
  url		= {https://onlinelibrary.wiley.com/doi/abs/10.1002/jcc.10043},
  year		= {2002}
}

@Article{	  Dobson:Alternative.1993,
  author	= {Dobson, John F.},
  title		= {Alternative expressions for the Fermi hole curvature},
  journal	= {J. Chem. Phys.},
  volume	= {98},
  number	= {11},
  pages		= {8870-8872},
  year		= {1993},
  month		= {06},
  issn		= {0021-9606},
  doi		= {10.1063/1.464444},
  url		= {https://doi.org/10.1063/1.464444}
}

@Article{	  Harada.Nakano.ea:Circularly.2009A,
  author	= {Harada, Takashi and Nakano, Yoko and Fujiki, Michiya and
		  Naito, Masanobu and Kawai, Tsuyoshi and Hasegawa,
		  Yasuchika},
  title		= {Circularly Polarized Luminescence of Eu(III) Complexes
		  with Point- and Axis-Chiral Ligands Dependent on
		  Coordination Structures},
  journal	= {Inorg. Chem.},
  volume	= {48},
  number	= {23},
  pages		= {11242-11250},
  year		= {2009},
  month		= {12},

  issn		= {0020-1669},
  doi		= {10.1021/ic901663w},
  url		= {https://doi.org/10.1021/ic901663w}
}

@Article{	  Tellgren.Soncini.ea:Nonperturbative.2008,
  author	= {Tellgren, Erik I. and Soncini, Alessandro and Helgaker,
		  Trygve},
  title		= {Nonperturbative ab initio calculations in strong magnetic
		  fields using London orbitals},
  journal	= {J. Chem. Phys.},
  volume	= {129},
  number	= {15},
  pages		= {154114},
  year		= {2008},
  month		= {10},
  issn		= {0021-9606},
  doi		= {10.1063/1.2996525},
  url		= {https://doi.org/10.1063/1.2996525}
}

@Article{	  Soncini.Teale.ea:Maps.2008,
  author	= {Soncini, A. and Teale, A. M. and Helgaker, T. and De
		  Proft, F. and Tozer, D. J.},
  title		= {Maps of current density using density-functional methods},
  journal	= {J. Chem. Phys.},
  volume	= {129},
  number	= {7},
  pages		= {074101},
  year		= {2008},
  month		= {08},
  issn		= {0021-9606},
  doi		= {10.1063/1.2969104},
  url		= {https://doi.org/10.1063/1.2969104},
}
\end{document}